\documentclass[fleqn,usenatbib]{rasti}

\usepackage{newtxtext,newtxmath}

\usepackage[T1]{fontenc}
\usepackage{ae,aecompl}

\usepackage{hyperref}
\usepackage{float}

\usepackage{bm}
\usepackage[usenames, dvipsnames]{color}
\usepackage{caption}
\usepackage{subcaption}
\usepackage{graphicx}	
\usepackage{amsmath}	
\usepackage{amssymb}	

\graphicspath{{./Plots/}}

\usepackage{tikz,xcolor,hyperref}

\definecolor{lime}{HTML}{A6CE39}
\DeclareRobustCommand{\orcidicon}{
    \begin{tikzpicture}
    \draw[lime, fill=lime] (0,0)
    circle [radius=0.16]
    node[white] {{\fontfamily{qag}\selectfont \tiny ID}};
    \draw[white, fill=white] (-0.0625,0.095)
    circle [radius=0.007];
    \end{tikzpicture}
    \hspace{-2mm}
}

\foreach \x in {A, ..., Z}{\expandafter\xdef\csname orcid\x\endcsname{\noexpand\href{https://orcid.org/\csname orcidauthor\x\endcsname}
            {\noexpand\orcidicon}}
}

\title[Overcoming Separation Between Counterparts Due to Unknown Proper Motions]{Overcoming Separation Between Counterparts Due to Unknown Proper Motions in Catalogue Cross-Matching}

\author[Tom J. Wilson]{
Tom J. Wilson$^{1}$\thanks{Email: t.j.wilson@exeter.ac.uk; onoddil@pm.me}\orcidA{}
\\
$^{1}$School of Physics, University of Exeter, Stocker Road, Exeter EX4 4QL, UK\\
}

\date{Accepted XXX. Received YYY; in original form ZZZ}

\pubyear{2022}

\begin{document}
\label{firstpage}
\pagerange{\pageref{firstpage}--\pageref{lastpage}}
\maketitle
\begin{abstract}
To perform precise and accurate photometric catalogue cross-matches -- assigning counterparts between two separate datasets -- we need to describe all possible sources of uncertainty in object position.
With ever-increasing time baselines between observations, like 2MASS in 2001 and the next generation of surveys, such as the Vera C. Rubin Observatory's LSST, \textit{Euclid}, and the \textit{Nancy Grace Roman} telescope, it is crucial that we can robustly describe and model the effects of stellar motions on source positions in photometric catalogues.
While \textit{Gaia} has revolutionised astronomy with its high-precision astrometry, it will only provide motions for $\approx$10\% of LSST sources; additionally, LSST itself will not be able to provide high-quality motion information for sources below its single-visit depth, and other surveys may measure no motions at all.
This leaves large numbers of objects with potentially significant positional drifts that may incorrectly lead matching algorithms to deem two detections too far separated on the sky to be counterparts.

To overcome this, in this paper we describe a model for the statistical distribution of on-sky motions of sources of given sky coordinates and brightness, allowing for the cross-match process to take into account this extra potential separation between Galactic sources.
We further detail how to fold these probabilistic proper motions into Bayesian cross-matching frameworks, such as those of Wilson \& Naylor.
This will vastly improve the recovery of e.g. very red objects across optical-infrared matches, and decrease the false match rate of photometric catalogue counterpart assignment.

\end{abstract}


\section{Introduction}
Counterpart assignment, the merging of bandpass detections in two (or more) datasets, enables a wide range of value-added science, and is therefore a crucial aspect of many areas of astronomical research.
Fundamentally, we require the ability to answer the question `are these two detections observations of two different objects, or two observations or the same astrophysical object?'
Thus, to provide accurate and precise cross-matches between two photometric catalogues, we require a complete description of all sources of separation between detections of a single astrophysical object.

Unfortunately for astronomers, there are many reasons for the same source, detected by two different telescopes in different parts of the world at different times, to have recorded positions that are not perfectly aligned with one another.
The first, and frequently assumed only, contribution is that from the act of measuring the position of the source on the detector image as part of the catalogue creation process.
This `centroid' uncertainty is related to the size of the telescope and the wavelength of the observation, as well as the atmospheric seeing, if applicable -- all of which affect the telescope `point spread function' (PSF), as well as the signal-to-noise ratio (SNR) of the detection, related to its brightness (e.g. \citealp{King:1983aa}).
\citet{2017MNRAS.468.2517W} highlighted an additional source of positional shift that can affect detections in crowded fields: sources, too close together on the sky to be resolved by the telescope, can appear as a single observation, leading the fainter source to influence the position of the (assumed singular) brighter object (sometimes referred to as `classical confusion'; see also e.g. \citealp{2001AJ....121.1207H}).

Here we consider an extra source of apparent separation between detections: that of the physical motion of the source across the sky.
If the observations to be combined are sufficiently separated in time, the `proper motion' of sources introduces a drift in the separation between consecutive measurements of the objects' locations.
Thus for pairs of observations with significant baselines, the proper motion-induced separations can become significant for large enough numbers of objects that, if we failed to consider these motions, we would decide the objects were too far apart to be counterpart detections of one object, and fail to assign them properly.

For probabilistic cross-matching algorithms this problem of object drift is further compounded.
It is not only objects with significant proper motion that suffer, those few objects with motions large enough to render them completely incompatible with the hypothesis that the two detections are counterparts.
All objects, even those with relatively small motions, are affected.
Any motion on the same length scale as the astrometric precisions will impact the derived match confidence, and potentially render quoted match or non-match probabilities meaningless.
This issue of your chosen model completely encapsulating the information contained within your data (or not) is often referred to as `model (mis)specification'.
Therefore, the effect of source motion must be accounted for, even if not so extreme as to completely move an object beyond its prior position.
If it is not taken into account, users of any resulting cross-match tables may not be able to put trust in the quoted match likelihoods and be able to take reliable, high-confidence cuts of the merged datasets.

This effect is particularly important for the upcoming Vera C. Rubin Observatory's Legacy Survey of Space and Time (LSST; \citealp{Ivezic2019}), for a few key reasons.
First, it will operate from $\sim$2025-2035, and thus have a two or three-decade baseline to the numerous surveys that operated during the 2000s and 2010s, such as 2MASS \citep{Skrutskie:2006um} or SDSS (e.g. \citealp{York2000}).
And second, it will lack measured proper motions for almost all of its sources for a large fraction of its survey lifetime.
In part this is because the survey will require a multi-year baseline before reliable proper motions can be derived, but more simply because most objects within the full LSST catalogue will be below the completeness limit of the single-visit images.
Even in this specific case, with Rubin's high-fidelity time-series capabilities, proper motions will only ever be available for objects that appear in multiple images, which sets the proper motion magnitude limit much higher than that of inclusion in the full coadd catalogue.
Worse still, the sheer number density of objects in the full LSST catalogue mean that up to 10 LSST sources will be potential counterparts to every single opposing catalogue object, and the `re-shuffle' of objects, even of order the precision of the measured positions, may lead to false matches being returned for a sizeable fraction of the catalogues.
Thus, the survey will be especially susceptible to this match misspecification due to proper motion drift, primarily at faint magnitudes.
Additionally, other upcoming missions such as \textit{Euclid} \citep{Laureijs2011} and the \textit{Nancy Grace Roman} Space Telescope \citep{Green2012} will likely lack the multi-epoch capabilities that Rubin and LSST offer, but still suffer the effects of decade-long time baselines back to previous generations of deep surveys, such as SDSS or VISTA (e.g. VHS, \citealp{McMahon2013}).

As time goes on, and we accumulate increasing numbers of surveys we wish to combine to maximum scientific return, we will increasingly no longer be able to ignore even relatively small levels of apparent on-sky motion.
One obvious solution is to use the \textit{individual} proper motions available through datasets such as the \textit{Gaia} \citep{2016A&A...595A...1G} mission.
These positions -- combined with the rate-of-change of position from the proper motions -- can be `fast-forwarded' through time, allowing for sources to be placed in the epoch of the opposing catalogue, removing on-sky drift as a factor in considering the separation between sources.
However, this is impractical for surveys such as LSST for a couple of reasons.

First, and simplest, is that proper motions are not available for the entire \textit{Gaia} catalogue.
Something like 20\% of sources in the early Data Release 3 (eDR3; \citealp{Collaboration2021}; \citealp{Lindegren2021}) do not have the five- or six-parameter solutions necessary to include proper motions, and a not insignificant fraction of those that have quoted proper motions have uncertainties that render the quoted values useless for any meaningful position projection.
Second, this would result in needing to run two \textit{Gaia}-to-other-catalogue matches, merge the most likely of those matches in turn, and then run an internal \textit{Gaia}-\textit{Gaia} look up to get the inner join of the two catalogues.
This would significantly affect the quality of the resulting matched datasets, with probabilistic cross-matching processes not being able to provide the proper probability of sources in the two `other catalogue' datasets matching.
Third, and most crucial, is the dynamic range consideration.
\textit{Gaia} is, for all its superb data, a relatively bright survey -- at least by LSST standards.
It also lacks coverage against longer wavelength surveys, where Galactic extinction is less oppressive.
LSST will, with its $\sim$7 magnitude fainter completeness limit, include at least an order of magnitude more stars \citep{Ivezic2019}, and VISTA, as an example infrared (IR) catalogue of consideration as an ancillary dataset to extend LSST information with, will have little overlap with \textit{Gaia} due to differing wavelength coverage.
Other surveys with deeper completeness limits, such as \textit{Euclid}, \textit{Roman}, and SDSS, will also suffer significant numbers of matches beyond the \textit{Gaia} completeness limit, with neither offering reliable proper motions at 21st magnitude or fainter.
Thus, even if we did decide to peg \textit{Gaia} as our gold standard, this would leave perhaps 9 out of every 10 LSST Galactic sources without a proper motion match.
Those LSST objects, and many others in other catalogues, would be in need of a separate, and worse, cross-match once we had handled those few objects with a \textit{Gaia} proper motion.

If we cannot trust matches between observations with significant time between observations, and we cannot necessarily use ancillary datasets with measured proper motions, how can we recover robust catalogue counterpart assignments through cross-matching?
Naively, we might think that we can `re-center' the distribution of offsets, by subtracting some mean separation between all of our counterpart pairings to account for the drift of our objects.
However, different average proper motions across the dynamic range of the two catalogues would cause further systematics, affecting the distribution of counterpart separations.
One may also think to use a Galactic model that calculates stellar velocities, and hence provides proper motions as viewed from Earth, for example the Besan\c con model \citep{Robin2003}.
However, as discussed in more detail in Section \ref{sec:besanconcomparison}, there are various reasons that these proper motions do not provide robust enough statistics for the determination of a statistical separation drift between two potential counterpart stars during a probabilistic cross-match process.

Thus, in this paper we put forward a model to build a \textit{statistical} distribution of proper motions of sources, based on their Galactic motions.
Modelling all sources of motion a star orbiting the Galactic center is subject to -- its bulk circular orbit, and any `random' scatter of sources from e.g. stellar cluster interactions -- combined with the Sun's motion, we are able to build a picture of the apparent motion of the given object.
To do so, we begin with the velocity of the star as it orbits around the Galactic center.
The conversion from velocity -- in units like $\mathrm{km}\,\mathrm{s}^{-1}$ -- to proper motion -- in units like $\mathrm{arcsec}\,\mathrm{yr}^{-1}$ -- is, roughly speaking, an inverse relation with distance.
As we detail later, instead of using distance directly, we intend to use the brightness of an object as its more readily available surrogate, accepting that this is only an approximation.

Hence, combining the apparent motions of a wide range of objects at the same sky position and brightness we obtain all proper motions such a source might have -- faint M dwarfs close by would have larger apparent motions than very intrinsically bright supergiants, but all `types' of object contribute to the spread of motion drifts a source of this particular brightness could have.
We are not particularly interested in a precise reconstruction of Galactic orbital dynamics -- not being overly concerned with the details of spiral arm dynamics, or the specifics of the Milky Way Bar shape, for example.
The use of the proper motions here is to `spread' the motion drift, the separation between potential counterparts, allowing for the recovery, and increasing the reliability of match probability, of these objects within a Bayesian cross-matching framework.
Therefore, the exact shape is less important than its central location and width; so long as these match reality to a fraction of the precision of the objects' positions and the underlying distribution width to a factor $1.5-2$, the model has served its purpose.
The probability of two stars being counterpart ranges over many orders of magnitude, and hence the resulting probability density functions (PDFs) we derived to model the statistical proper motions having widths, and overall PDF heights, correct to a factor two is sufficient to improve counterpart recovery.
We also desire computational simplicity, and thus speed, over an overly prescriptive or detailed exact model of the Galaxy, as this model must fit within a wider counterpart assignment framework and be able to be run on-the-fly for some arbitrary sets of sky positions and brightnesses.

Once we have constructed our distribution of theoretical proper motions, we can then consider their effect on our potential cross-match pairings.
Here we can, in a similar way to how we would handle known \textit{Gaia} proper motions, translate one object's position into the epoch of the second catalogue observations, and consider the additional separation caused by the motion of the object.
We must then test all modelled proper motions, with appropriate weighting, ultimately accounting for these potential additional time-based separations in answering the question of whether these two detections are two physical sky objects or one source viewed twice in time.

\subsection{Paper Layout}
This paper is split into two parts.
First, we detail the construction of a simple analytic model for the statistical distribution of potential proper motions of a source of a given magnitude and sky coordinates.
In Section \ref{sec:constructpms} we detail the constituent parts necessary to build the model of proper motions.
We describe the process of building the distributions in Section \ref{sec:createpmdist}, while Section \ref{sec:assesspmmodel} discusses the precision and accuracy of the model at various Galactic sightlines and brightnesses.
Here we will return to \textit{Gaia}, using its high-precision stellar proper motions across many Galactic sightlines to evaluate and corroborate our model distributions.

The second part of this work describes how to include this unknown proper motion distribution -- or any distribution of proper motions, theoretical or poorly constrained yet detected -- in the cross-matching process.
We discuss the mathematical framework necessary for including proper motion drift in the evaluation of the separation between potential counterpart detections in Section \ref{sec:includeinmatching}.
We also touch upon how to use these statistical distributions of proper motion drift as a discriminator between stars and galaxies.
Concluding remarks are given in Section \ref{sec:conclusions}.

In Appendix \ref{sec:coordsystems} we outline the various coordinate systems and derive the transformation matrices used throughout this work, while in Appendix \ref{sec:convolutionmaths} we detail the inclusion of the proper motion PDFs within a probabilistic cross-match astrometric separation likelihood framework.

\section{Constructing the Proper Motions}
\label{sec:constructpms}
We need to build a model to describe the observed motions of sources across the sky.
There are, essentially, three components that matter: first, the `peculiar' motion of the Sun itself; second, the expected velocity of a source moving with the Galactic rotation; and third, the random velocities of the Galactic sources; these will be discussed individually.
First, we must consider how we will build this model.

As the model involves consideration of the Galactic rotation (and we will see later that source random motion is location dependent), we will require a description of the position of the source in the Galaxy.
The first two components are easy: sky coordinate in Galactic coordinates (converting Equatorial $\alpha$ and $\delta$ by rotation to longitude $l$ and latitude $b$, if necessary).
The only other component we would need is a distance, or parallax; however, if we have parallax we likely have a unique proper motion, as these are generally fit for simultaneously, and hence we use the next best proxy: magnitude.
This will blend several `types' of source together (e.g. dwarfs and giants of the same brightness are at different distances).
We will see that while this might introduce extra scatter in the proper motion drifts, our models match \textit{Gaia} proper motions at magnitude cuts well -- and indeed account for the fact that we do not know the type of any individual source from its photometry \textit{a priori}!

However, we do need \textit{some} distance metric, and hence we turn to the TRILEGAL simulations \citep{Girardi2005} to provide a theoretical magnitude-distance relation.
Thus, while our catalogue sources have their proper motions built as a function of sky coordinates and photometric brightness, our model is coordinate/distance based.
In the following sections we describe how we formulate a description of the observed proper motion of a set of sources based on their given parameters.

\subsection{Solar Peculiar Motion}
The first component in the Galactic motion is the unique velocity of the Sun, relative to the local standard of rest (LSR).
The Sun's motion through the Galaxy will induce a `secular' parallax effect (i.e. distance-dependent, albeit non-periodic, as a trigonometric parallax would be) in the apparent movement of all other sources in the sky, with opposite sign.
Hence we need to know the Sun's motion, relative to this `zero point' motion, the LSR, defined in the Heliocentric Cartesian coordinate frame, $(U,\,V,\,W)$ -- velocities corresponding to the $(x,\,y,\,z)$ coordinate system.
Here we use the values of \cite{Schoenrich2010}:
\begin{align}
\begin{split}
    U_\odot = 11.1\,\mathrm{km}\,\mathrm{s}^{-1}\\
    V_\odot = 12.2\,\mathrm{km}\,\mathrm{s}^{-1}\\
    W_\odot = 7.3\,\mathrm{km}\,\mathrm{s}^{-1}.
\end{split}
\end{align}

\subsection{Galactic Rotation}
The main component of our model that will dictate the Galaxy-wide observed motions of sources is that of the Galactic rotation, and the stellar streaming.
Descriptions of this motion go back to \citet{Oort1927}, with the Oort constants describing the motion of stars on closed orbits around the Galaxy.
However, through modern kinematic surveys, obtaining the three-dimensional velocities and independent distances to a host of well-characterized objects, it is possible to directly measure the rotation curve of the Milky Way.
Thus, we can derive the average tangential velocity of sources orbiting the Galactic center at a given radius, here following Model 3 of \citet{Mroz2019}.

Obtaining the rotational velocity $\Theta$ at a given Galactocentric radius $R_c$, we can transform this Galactocentric Cylindrical azimuthal velocity into a Galactocentric Cartesian coordinate frame and subtract the Solar peculiar and LSR motion, obtaining $U_1$, $V_1$, and $W_1$ (\citealp{Mroz2019}, equations 5-7).
To do so, we use the transformation
\begin{equation}
\bm{\mathcal{T}}_t = \left(\begin{matrix} \frac{R_c^2 + R_\odot^2 - d_\mathrm{ip}^2}{2\,R_c\,R_\odot} & \frac{d_\mathrm{ip}}{R_c} \sin(l) & 0 \\ -\frac{d_\mathrm{ip}}{R_c} \sin(l) & \frac{R_c^2 + R_\odot^2 - d_\mathrm{ip}^2}{2\,R_c\,R_\odot} & 0 \\ 0 & 0 & 1 \end{matrix}\right);
\label{eq:cylcart}
\end{equation}
here $R_\odot$ is the Solar Galactocentric Cylindrical radius, $d_\mathrm{ip}$ is the in-plane distance from the Sun to the particular location, and $l$ is Galactic longitude -- see Appendix \ref{sec:cylcartrot} for details.
Deviating from \citet{Mroz2019}, we set $(U_s,\,V_s,\,W_s)$, the non-circular motion of the source, all to zero, and hence have
\begin{align}
\begin{split}
    U_1 &= \Theta(R_c) \times \frac{d_\mathrm{ip}}{R_c}\sin(l) - U_\odot\\
    V_1 &= \Theta(R_c) \times \frac{R_c^2 + R_\odot^2 - d_\mathrm{ip}^2}{2\,R_c\,R_\odot} - V_\odot - \Theta_\odot\\
    W_1 &= -W_\odot.
    \label{eq:uvw1}
\end{split}
\end{align}
Once we have the Galactocentric Cartesian components of the rotational velocity, relative to the Sun, we can transpose into the in-plane Heliocentric radial and tangential velocities,
\begin{align}
\begin{split}
    v_d &= U_1 \cos(l) + V_1 \sin(l)\\
    v_l &= -U_1 \sin(l) + V_1 \cos(l),
\end{split}
\end{align}
along with $v_z = W_1$, since the two axes are still in alignment.
Finally, once we have appropriate Heliocentric Cylindrical velocities, we can construct our latitudinal velocity as
\begin{equation}
    v_b = v_z \cos(b) - v_d \sin(b)
\end{equation}
and relate the Heliocentric longitudinal and latitudinal velocities to their proper motions through
\begin{align}
     \mu_{l*} \equiv \mu_l\cos(b) &= k \times \pi v_l, \label{eq:muldef}\\
     \mu_b &= k \times \pi v_b,
     \label{eq:mubdef}
\end{align}
where $\pi$ is the parallax of the source.
As our distances come from Galactic models, we assume they are not subject to any observational bias or uncertainty, and simply treat $\pi^{-1} = d$.
The factor $k$ describes the translation from units of $\mathrm{km}\,\mathrm{s}^{-1}\,\mathrm{kpc}^{-1}$ to $\mathrm{mas}\,\mathrm{yr}^{-1}$, and is given by $k = 0.2108\,\mathrm{mas}\,\mathrm{yr}^{-1}\,\mathrm{km}^{-1}\,\mathrm{s}\,\mathrm{kpc}$.

In practice, the TRILEGAL simulations do not provide either distance or parallax, but provide its distance modulus.
Hence, to obtain a (three-dimensional) distance $d$ in kpc, we invert the absolute magnitude equation:
\begin{equation}
    d = 10^{-3} \times 10^{0.2(m - M) + 1}
\end{equation}
where $m - M$ is the distance modulus.

\subsection{Asymmetric Drift Velocity}
\label{sec:asymdrift}
The above equations describe the \textit{average} Galactic rotation velocity around the center of the Galaxy.
However, objects have other sources of velocity that impact their observed proper motions, and this leads to a deviation away from the expected velocity, and thus proper motion.
This is termed the asymmetric drift velocity, and essentially controls how much of the theoretical velocity a source \textit{should} have is taken by other, random motions.
Thus, we must include this component in the velocities.

We model three Galactic components (see Sections \ref{sec:thindiskmodel}-\ref{sec:haloconstruction} for more details of their construction) in our simulations: the Galactic thin and thick discs, and a single Galactic (outer) halo.
Each of these is given their own asymmetric drift, as a measure of the levels to which their motions are different from `pure' streaming motion.
We assume the thin disc of the Galaxy has an azimuthal asymmetric drift velocity of $10\,\mathrm{km}\,\mathrm{s}^{-1}$ \citep{Robin2003}.
As \citet{Mroz2019} used Classical Cepheids in the derivation of their rotation curve, we also assume the rotation curve is valid for the thin disc and already folds in any drift velocity, and therefore just need to consider the \textit{relative} drift velocities of the thick disc and the halo.
We use a thick disc drift velocity of $49\,\mathrm{km}\,\mathrm{s}^{-1}$ \citep{Pasetto2012a}, and give the halo a drift velocity of $240\,\mathrm{km}\,\mathrm{s}^{-1}$ \citep{Golubov2013}, to essentially counteract the motion of the LSR, modelling the halo as stationary relative to the Galaxy.
We therefore use $v_{a, \phi} = \{0\,,39\,,230\}\,\mathrm{km}\,\mathrm{s}^{-1}$ for the relative thin disc, thick disc, and halo drift velocities, respectively.

For a given location in the Galaxy, our decomposition of the drift velocity, from Galactocentric Cylindrical coordinates into Heliocentric Cylindrical coordinates, is given by the transformation
\begin{equation}
    \bm{v}{'}_\mathrm{\!\!drift} = \bm{\mathcal{T}}_c\,\bm{v}_\mathrm{drift},
\end{equation}
with
\begin{equation}
\bm{\mathcal{T}}_c = \left(\begin{matrix} \frac{R_c^2 + d_\mathrm{ip}^2 - R_\odot^2}{2R_cd_\mathrm{ip}} & \frac{R_\odot}{R_c} \sin(l) & 0 \\ \frac{R_\odot}{R_c} \sin(l) & -\frac{R_c^2 + d_\mathrm{ip}^2 - R_\odot^2}{2R_cd_\mathrm{ip}} & 0 \\ 0 & 0 & 1 \end{matrix}\right)
\label{eq:cylrot}
\end{equation}
and
\begin{equation}
\bm{v}_\mathrm{drift} = \left(\begin{matrix} 0 \\ v_{a,\phi} \\ 0 \end{matrix}\right),
\end{equation}
with $v_{a,\phi}$ taking on any one of the three given drift velocities above, depending on which component of the Galaxy is being considered. For details on the derivation of this transformation (rotation and mirror) matrix, see Appendix \ref{sec:cyl_to_cyl_rot}.

\subsection{Velocity Dispersion}
\label{sec:veldisp}
The asymmetric drift velocity suggests that some of the motion that ought to be used by a given source in its rotation around the Galaxy is being used otherwise, in a random component.
Hence, a collection of sources in a particular part of the Galaxy will have some spread of their velocities around some mean value.
Thus, to be able to model our collection of sources in the Galaxy we require a description of the dispersion of the velocities.

Each component of the Galaxy modelled -- thin and thick discs, and halo -- have their own prescription of velocity dispersion.
In addition, when simulating sources, we do not initially know to which component to assign a given simulated object.
Therefore we simply generate a set of proper motion realisations for each of the three components, weighted according to their \textit{a priori} density at that location.
Hence, in the following sections we also describe the formulation of each component's density profile, which are simply re-normalised by the sum of their densities to provide a prior probability, used as the weight for the proper motion distribution.
We currently do not consider the Bulge, a common component of Galactic simulation models such as TRILEGAL, in our proper motion model.
We chose to ignore this additional component for this initial, exploratory model, focussing on relatively bright sources, with detected \textit{Gaia} proper motions to compare and verify our model against.
This should mean they are sufficiently far from the Galactic center to not be influenced by the Bulge, avoiding the complexities that the inner region of the Galaxy impose on velocities of orbiting stars.
However, with the upcoming LSST survey and the need to model much fainter, more distant objects in the next few years, we will investigate a robust Galactic Bulge/Bar model for inclusion within this proper motion framework.
We simply conclude, for now, that it would be easy to add additional components, and we could model a simple Bulge component after e.g. \citet{Jackson2002}.

Once a given Galactic component has its dispersion vector -- its covariance matrix -- in the Heliocentric Cylindrical coordinate system, then a realisation is drawn from a multivariate normal, given by
\begin{equation}
    \bm{v}_{\mathrm{noisy}, i} \sim \mathcal{N}(\bm{v}- \bm{v}{'}_{\mathrm{\!\!drift}, i}, \bm{\Sigma}{'}_{\!i})
\end{equation}
where $i \in \{\mathrm{thin}, \mathrm{thick}, \mathrm{halo}\}$ and $\bm{\Sigma}{'}_{\!i}$ is the Cylindrical frame rotated covariance matrix of the $i$th component.

\subsubsection{Thin Disc}
\label{sec:thindiskmodel}
The thin disc is modelled as an exponentially decaying density profile with given radial and vertical scale heights, as per \citet{Juric2008} and \citet{Ivezic2008}:
\begin{equation}
    \rho(R_c, z) = \Gamma \exp\left(-\frac{R_c - R_\odot}{l_\mathrm{thin}} - \frac{z + z_\odot}{h_\mathrm{thin}}\right),
\end{equation}
with $R_\odot = 8.09\,\mathrm{kpc}$ \citep{Mroz2019}, $z_\odot = 25\,\mathrm{pc}$ \citep{Juric2008}; $\Gamma$ is an irrelevant normalisation constant, used simply to explicitly cancel in the re-normalisation from density to weighting.
The radial and vertical scale lengths we use are the bias-corrected values calculated by \citeauthor{Juric2008}: $l_\mathrm{thin} = 2.6\,\mathrm{kpc}$, and $h_\mathrm{thin} = 0.3\,\mathrm{kpc}$.

The dispersion vector for the thin disc is based on observations of RAVE stars from \citet{Pasetto2012}.
However, the nature of the observations limit their calculation of covariances to approximately $1\,\mathrm{kpc}$ from the Sun, and we need to extrapolate these dispersions out to perhaps five times that distance.
Thus we turn to \citet{Amendt1991} for relations between the various (co-)variances in the dispersion vector.

First, we assume that the variance in the vertical direction, $\sigma_{zz}^2$, scales with \textit{radial} distance in the mid-plane of the Galaxy:
\begin{align}
\begin{split}
    \sigma_{zz}^2(R_c, z=0) &= \sigma_{zz}^2(0, 0) \exp\left(-\frac{R_c}{l_\mathrm{thin}}\right) \\ &= \sigma_{zz}^2(R{'}_{\!\!\odot}, 0) \exp\left(-\frac{R_c - R{'}_{\!\!\odot}}{l_\mathrm{thin}}\right).
\end{split}
\end{align}
As discussed by \citeauthor{Amendt1991}, this scaling relation is also sometimes assumed for $\sigma_{R_cR_c}^2$, but a second, valid formalism can be used where the rotation curve of the Galaxy is flat -- which it should be safe to assume given the very small gradient from \citet{Mroz2019} for most of the Galaxy.
This formalism is based on a constant \citet{Toomre1964} local stability parameter, and gives
\begin{equation}
    \sigma_{rr}^2 \equiv \sigma_{R_cR_c}^2 \propto R_c^2 \exp\left(-2\frac{R_c}{h_\mathrm{thin}}\right).
\end{equation}
Hence we use\footnote{Using $\sigma_{rr}^2$ for the Galactocentric Cylindrical frame radial dispersion component, to avoid the slightly clunky notation $\sigma_{R_cR_c}^2$.}
\begin{equation}
    \sigma_{rr}^2(R_c, 0) = \sigma_{rr}^2(R{'}_{\!\!\odot}, 0) \left(\frac{R_c}{R{'}_{\!\!\odot}}\right)^2 \exp\left(-2\frac{R_c - R{'}_{\!\!\odot}}{h_\mathrm{thin}}\right).
\end{equation}

For the vertical extrapolation, we assume a local Taylor expansion to first order (i.e. we extrapolate linearly to above and below the plane, from $z = 0$).
We limit this extrapolation to the inner one kiloparsec of the plane, and assume a constant dispersion beyond that, and hence:
\begin{align}
    \sigma_{zz}^2(R_c, z) &\simeq \sigma_{zz}^2(R_c, 0) + \min\left(1\,\mathrm{kpc},\,\lvert z \lvert\right) \frac{\partial \sigma_{zz}^2(R_c, 0)}{\partial \lvert z \lvert} \\
    \sigma_{rr}^2(R_c, z) &\simeq \sigma_{rr}^2(R_c, 0) + \min\left(1\,\mathrm{kpc},\,\lvert z \lvert\right) \frac{\partial \sigma_{rr}^2(R_c, 0)}{\partial \lvert z \lvert}.
\end{align}

Using the \citeauthor{Pasetto2012} data in the range $8.2\,\mathrm{kpc} \leq R_c \leq 8.8\,\mathrm{kpc}$, $-0.5\,\mathrm{kpc} \leq z \leq 0.5\,\mathrm{kpc}$, we find
\begin{align}
\begin{split}
\sigma_{zz}^2(R{'}_{\!\!\odot}, 0) &= 243.71\,\mathrm{km}^2\,\mathrm{s}^{-2},\\
\frac{\partial \sigma_{zz}^2(R_c, 0)}{\partial \lvert z \lvert} &= 306.84\,\mathrm{km}^2\,\mathrm{s}^{-2}\,\mathrm{kpc}^{-1},\\
\sigma_{rr}^2(R{'}_{\!\!\odot}, 0) &= 715.93\,\mathrm{km}^2\,\mathrm{s}^{-2},\\
\frac{\partial \sigma_{rr}^2(R_c, 0)}{\partial \lvert z \lvert} &= 1236.97\,\mathrm{km}^2\,\mathrm{s}^{-2}\,\mathrm{kpc}^{-1}.
\end{split}
\end{align}
Additionally, to be consistent with the data as derived by \citeauthor{Pasetto2012}, we use $R{'}_{\!\!\odot} = 8.5\,\mathrm{kpc}$ -- note that $R{'}_{\!\!\odot} \neq R_\odot$ -- for extrapolating the dispersions.
Here we have assumed their quoted location of the Sun `in the range $R \in\ ]8.4, 8.6]\,\mathrm{kpc}$' implies\footnote{Inclusive of 8.6$\,\mathrm{kpc}$ but exclusive of 8.4$\,\mathrm{kpc}$, equivalent to $(8.4, 8.6]$.} an assumed default location in the middle of the bin.

We assume, following \citet{Amendt1991} and \citet{Vallenari2006}, that
\begin{equation}
    \sigma_{rz}^2(R_c, z) \simeq \sigma_{rz}^2(R_c, 0) + z \frac{\partial \sigma_{rz}^2(R_c, 0)}{\partial z}
\end{equation}
where the first term on the right hand side vanishes by symmetry at $z=0$, and the derivative is given by
\begin{equation}
    \frac{\partial \sigma_{rz}^2(R_c, 0)}{\partial z} = \lambda(R)\frac{\sigma_{rr}^2(R_c, 0) - \sigma_{zz}^2(R_c, 0)}{R_c}.
\end{equation}
Given no information on the radial dependence of $\lambda$, we fix it to the local value of $\lambda = 0.6$ \citep{Amendt1991}.
In cases where the linear extrapolation would result in a correlation $\left(\rho_{rz} \equiv \frac{\sigma_{rz}^2}{\sigma_{rr}\sigma_{zz}}\right)$ larger in absolute value than one, we force the correlation back to either +1 or -1.

\begin{figure*}
    \centering
    \includegraphics[width=\textwidth]{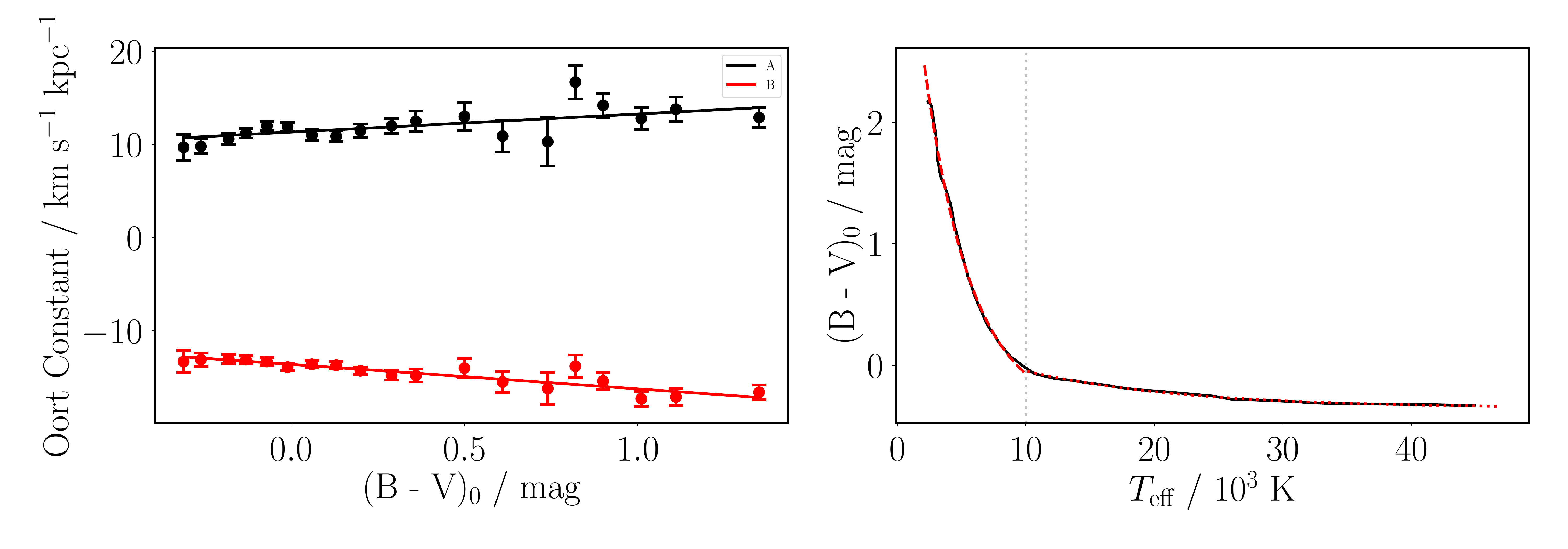}
    \caption{Relationships used to derive the dependencies of $A$ and $B$ on intrinsic colour.
             Left: linear relationships between intrinsic B-V and Oort constants, using the Oort constants as derived by \citet{Olling2003}.
             Right: a two-piece fit between effective temperature and intrinsic B-V, using the empirical colour sequence of \citet{Pecaut2013}.}
    \label{fig:colour_relation}
\end{figure*}

Following \citet{Vallenari2006}, this is the only off-diagonal term we consider for the thin disc covariance matrix.
Finally, again following the prescription of \citeauthor{Amendt1991}, we assume that the azimuthal and radial dispersions are related by a constant, and hence use
\begin{equation}
    \sigma_{\phi\phi}^2 = \frac{-B}{A - B} \sigma_{rr}^2,  
\end{equation}
with $A$ and $B$ the \citet{Oort1927} constants, for all $R_c$ and $z$.
We use the \citet{Olling2003} Oort constant values (their table 5, figure 6), as a function of intrinsic colour $B-V$.
Here we interpolate $A$ and $B$ as a linear function of $(B-V)_0$, fitting:
\begin{align}
\begin{split}
    A &= 1.94553 \times (B-V)_0 + 11.33138 \\
    B &= -2.63360 \times (B-V)_0 -13.60611
\end{split}
\end{align}
as shown in Figure \ref{fig:colour_relation} (left-hand panel).
As we are using TRILEGAL simulations, we require a conversion from available TRILEGAL parameters to $(B-V)_0$; for this we use the dwarf colour sequence of \citet{Pecaut2013}.
We fit a two-step function to the intrinsic B-V colour as a function of effective temperature (with $T$ in units of Kelvin),
\begin{align}
\begin{split}
    (B-V)_0 = \begin{cases}
        \begin{aligned}&-0.40739 + 5.07836\,\times \\&\exp(-0.27083 \times T/1000\mathrm{K}) \end{aligned}& T < 10000\,\mathrm{K}\\ \begin{aligned}&-0.35093 + 0.69012\,\times \\&\exp(-0.08179 \times T/1000\mathrm{K}) \end{aligned} & T \geq 10000\,\mathrm{K}\end{cases}
\end{split}
\end{align}
as shown in Figure \ref{fig:colour_relation}, right-hand panel.

Once we have all of the terms, we rotate the covariance matrix from its Galactocentric cylindrical coordinate frame into the Heliocentric coordinate system by
\begin{equation}
    \bm{\Sigma}{'} = \bm{\mathcal{T}}_c\,\bm{\Sigma}\,\bm{\mathcal{T}}_c^T = \bm{\mathcal{T}}_c\,\left(\begin{matrix} \sigma_{rr}^2 & 0 & \sigma_{rz}^2 \\ 0 & \sigma_{\phi\phi}^2 & 0 \\ \sigma_{rz}^2 & 0 & \sigma_{zz}^2 \end{matrix}\right)\,\bm{\mathcal{T}}_c^T
\end{equation}
using the rotation matrix as defined in equation \ref{eq:cylrot}.

\subsubsection{Thick Disc}
The formalism for the thick disc is very similar to that of the thin disc.
We also use the exponential decay model for the density profile, albeit with different scale lengths:
\begin{equation}
    \rho(R_c, z) = \Gamma f_\mathrm{thick}\exp\left(-\frac{R_c - R_\odot}{l_\mathrm{thick}} - \frac{z + z_\odot}{h_\mathrm{thick}}\right),
\end{equation}
again following the \citet{Juric2008} and \citet{Ivezic2008} formalism, with $R_\odot = 8.09\,\mathrm{kpc}$ and $z_\odot = 25\,\mathrm{pc}$ again, and $l_\mathrm{thick} = 3.6\,\mathrm{kpc}$, and $h_\mathrm{thick} = 0.9\,\mathrm{kpc}$.
In addition, the parameter $f_\mathrm{thick} = 0.13$ sets the relative densities of the thin and thick discs, and $\Gamma$ again is an arbitrary normalisation constant.

The thick disc dispersion vector uses the data from \citet{Pasetto2012a}, again following the same radial scaling relations for the thin disc:
\begin{align}
    \sigma_{rr}^2(R_c, 0) &= \sigma_{rr}^2(R{'}_{\!\!\odot}, 0) \left(\frac{R_c}{R{'}_{\!\!\odot}}\right)^2 \exp\left(-2\frac{R_c - R{'}_{\!\!\odot}}{h_\mathrm{thick}}\right) \\
    \sigma_{\phi\phi}^2(R_c, 0) &= \sigma_{\phi\phi}^2(R{'}_{\!\!\odot}, 0) \left(\frac{R_c}{R{'}_{\!\!\odot}}\right)^2 \exp\left(-2\frac{R_c - R{'}_{\!\!\odot}}{h_\mathrm{thick}}\right) \\
    \sigma_{zz}^2(R_c, 0) &= \sigma_{zz}^2(R{'}_{\!\!\odot}, 0) \exp\left(-\frac{R_c - R{'}_{\!\!\odot}}{h_\mathrm{thick}}\right)
\end{align}
where, once again, we use $R{'}_{\!\!\odot} = 8.5\,\mathrm{kpc}$ from \citet{Pasetto2012}, assuming the two papers were jointly analysed and hence have the same $R{'}_{\!\!\odot}$, although neither paper in the series quote a specific value.
This time, we do not describe any vertical dependency of the dispersions.
Finally, we take the diagonal terms as presented by \citet{Pasetto2012a} within or without the Solar circle as the values approximately at $(R{'}_{\!\!\odot}, 0)$, as given by their tables 3 and 4 respectively, but ignore the off-diagonal terms, which are all within $\approx1.5\sigma$ of zero.

Exactly the same as with the thin disc, we rotate the Galactocentric Cylindrical reference frame into Heliocentric Cylindrical coordinates by
\begin{equation}
    \bm{\Sigma}{'} = \bm{\mathcal{T}}_c\,\bm{\Sigma}\,\bm{\mathcal{T}}_c^T = \bm{\mathcal{T}}_c\,\left(\begin{matrix} \sigma_{rr}^2 & 0 & 0 \\ 0 & \sigma_{\phi\phi}^2 & 0 \\ 0 & 0 & \sigma_{zz}^2 \end{matrix}\right)\,\bm{\mathcal{T}}_c^T
\end{equation}
again using the cylindrical rotation matrix of equation \ref{eq:cylrot}.

\subsubsection{Halo}
\label{sec:haloconstruction}
The halo Galactic component follows the density profile
\begin{equation}
    \rho(R_c, z) = \Gamma f_h \left(\frac{R_\odot}{\sqrt{R_c^2 + \left(\frac{z}{q}\right)^2}}\right)^n,
\end{equation}
again using the \citet{Juric2008} and \citet{Ivezic2008} formalism, with $f_h = 0.0051$, $q = 0.64$, $n = 2.77$.
$\Gamma$ is a normalising constant once again.
This parameterization of an inverse power law leads, at $R_c = 0$, $z=0$, to an infinite halo density, and hence unphysical normalising weighting in the Galactic model.
We therefore truncate the halo density within the solar circle, $R_\odot$, fixing it at its value at $R_c = R_\odot$ at smaller radii.
This ought to be possible because the old Galactic halo should be negligible in relative density by the solar circle.

The dispersion vector for the halo is derived from \citet{King2015}, given in spherical coordinates.
We take the full covariance matrix from the closest radial bin from the `Equally Populated Bins' in their table 3 for a given set of $\left(R_s,\,\phi,\,\theta\right)$ parameters for a given source, with the exception of their $R_s = 12\,\mathrm{kpc}$ bin.
This bin gives a covariance matrix that is not positive semi-definite, and hence we ignore the off-diagonal terms for that individual bin.
These have no scaling applied to them and are taken exactly as quoted.

To rotate into the Heliocentric Cylindrical reference frame, we use
\begin{equation}
    \bm{\Sigma}{'} = \bm{\mathcal{R}}_{sc}\,\bm{\Sigma}\,\bm{\mathcal{R}}_{sc}^T
\end{equation}
where\footnote{Once again, $r$ has been used instead of $R_s$ for notation's sake, analogous to the thin and thick disc notations.}
\begin{align}
    \bm{\Sigma} &= \left(\begin{matrix} \sigma_{rr}^2 & \Sigma_{r\phi} & \Sigma_{r\theta} \\ \Sigma_{r\phi} & \sigma_{\phi\phi}^2 & \Sigma_{\phi\theta} \\ \Sigma_{r\theta} & \Sigma_{\phi\theta} & \sigma_{\theta\theta}^2 \end{matrix}\right), \\
    \bm{\mathcal{R}}_{sc} &= \bm{\mathcal{T}}_c \bm{\mathcal{R}}_s, \\
    \bm{\mathcal{R}}_s &= \left(\begin{matrix} \cos(\beta) & 0 & -d/R_s \sin(b)\ \\ 0 & 1 & 0 \\ d/R_s \sin(b) & 0 & \cos(\beta) \end{matrix}\right),
\end{align}
and where $\Sigma_{r\theta} \equiv \sigma_{r\theta}^2$, following the \citeauthor{King2015} notation.
$\bm{\mathcal{R}}_s$ describes the rotation from Galactocentric Spherical coordinates to Galactocentric Cylindrical coordinates, with $\bm{\mathcal{T}}_c$, as before, defining the rotation from Galactocentric Cylindrical to Heliocentric Cylindrical coordinates.
$\beta$ is defined as the angle between the spherical radial vector and the Galactic plane ($b = 0^\circ$), with $d$ the three-dimensional distance to the source in question, and $R_s$ the three-dimensional Galactocentric distance to the star.
For more details on the derivation of this transformation matrix, see Appendix \ref{sec:sph_to_cyl_rot}.

\section{Creating a Proper Motion Distribution}
\label{sec:createpmdist}
Now that we have described the model for simulating the velocity of a source at a given position in the Galaxy, we can create a theoretical distribution of sources.
In a small sky coordinate window (in our tests limiting ourselves to a few square degrees in the Galactic plane, and relatively small polar cap latitude windows), we run a TRILEGAL simulation in the center of the defined region.
We simulate either 1.5 million sources down to \textit{Gaia} $G = 25$, or as many as we are allowed within 10 square degrees, the maximum limit of the public simulation API.
Distances for these simulated sources are derived from their absolute distance modulus, and -- with no positional information in the simulated dataset -- we randomly place the sources within the rectangle defining the coordinate window.

For a given small magnitude range of sources, each source then has its proper motions calculated as though it were from each of the three Galactic components in turn, with some number of realisations ($N \approx 1000$) of the multivariate dispersion drawn.
We then calculate a weighted histogram of proper motions.
For each source, $j = 1, 2, ..., M$, where $M$ is the number of simulated stars (and thus distances), the three Galactic components at the given Galactic longitude, latitude, and distance have their respective weights $w_{ij}$ ($i \in \{1, 2, 3\}$, or $i \in \{\mathrm{thin}, \mathrm{thick}, \mathrm{halo}\}$) calculated.
The weighted histogram is therefore built with each derived proper motion being given weight $w_{ij}/N$ ($N$ the number of derived Galactic velocities for the $j$th source, in each of the three components), for each of the $3\times N \times M$ derived proper motions, across all $M$ objects.
The histogram (which will contain $M$ weighted counts across all bins) is then converted to a PDF.

\begin{figure}
    \centering
    \includegraphics[width=\columnwidth]{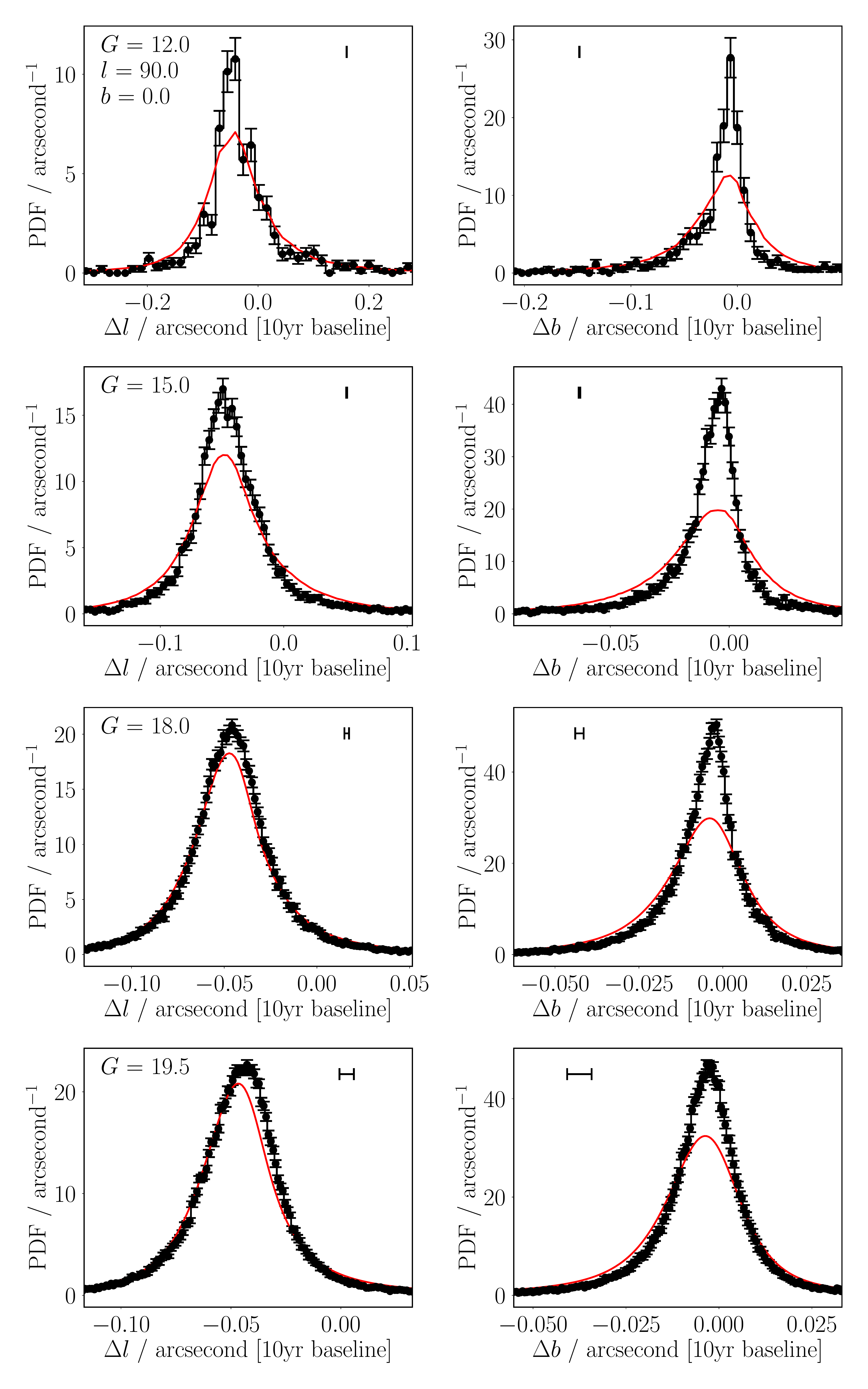}
    \caption{Distributions of proper motions (Galactic longitude, left-hand columns, and latitude, right-hand columns) for sources at $l = 90^\circ$,  $b = 0^\circ$, for $G=12$, $G=15$, $G=18$, and $G=19.5$ (each respective row).
             Proper motions have been converted from a per-year drift to decadal positional change.
             \textit{Gaia} proper motions are shown in the black histogram, with simulated distributions of proper motions in the red solid lines.
             Errorbars in the corner of each subplot show the typical uncertainty of each individual \textit{Gaia} proper motion, while the plot labels show the Galactic longitude and latitude, and $G$ magnitude, of the subset of sources.}
    \label{fig:good_90}
\end{figure}

For the purposes of visualisation and testing, we extract all of the proper motions of \textit{Gaia} eDR3 sources with flux SNRs greater than five in the same coordinate window and magnitude range defined for the simulated proper motions.
Finally, one additional step is taken, solely for the purposes of distribution comparison: we convolve the model with the median uncertainty of the \textit{Gaia} proper motions in the dataset for this magnitude cut and sightline.
This allows for the inclusion of non-negligible Gaussian uncertainties in our comparison of our generated model to the \textit{Gaia} data.
This step was purely for visualisation purposes, and is not part of the model itself.

\section{Assessing the Accuracy and Precision of the Proper Motion Model}
\label{sec:assesspmmodel}

\begin{figure}
    \centering
    \includegraphics[width=\columnwidth]{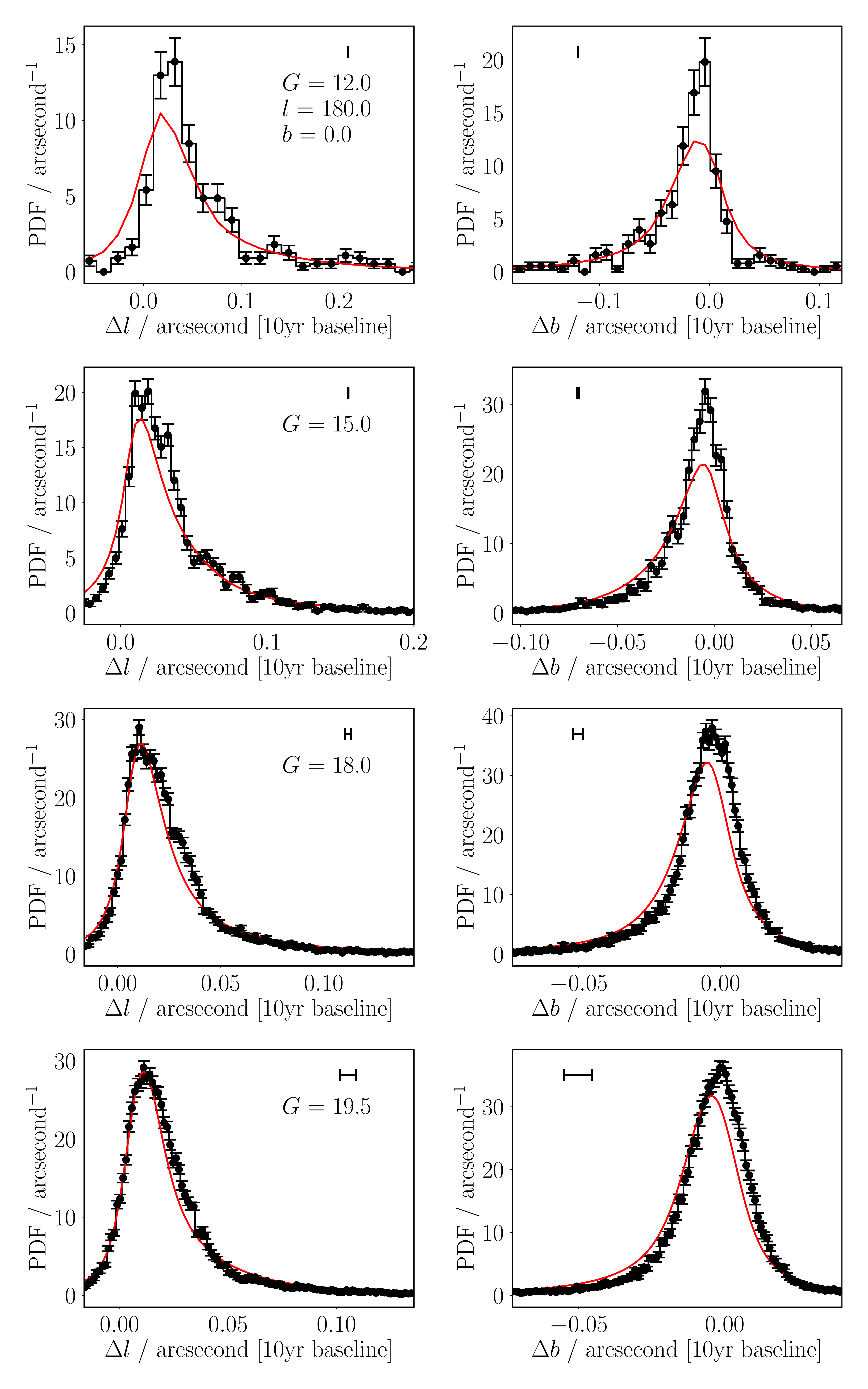}
    \caption{Distributions of proper motions for $l = 180^\circ$, $b=0^\circ$.
             Lines and symbols have the same meaning as in Figure \ref{fig:good_90}.}
    \label{fig:good_180}
\end{figure}

\begin{figure}
    \centering
    \includegraphics[width=\columnwidth]{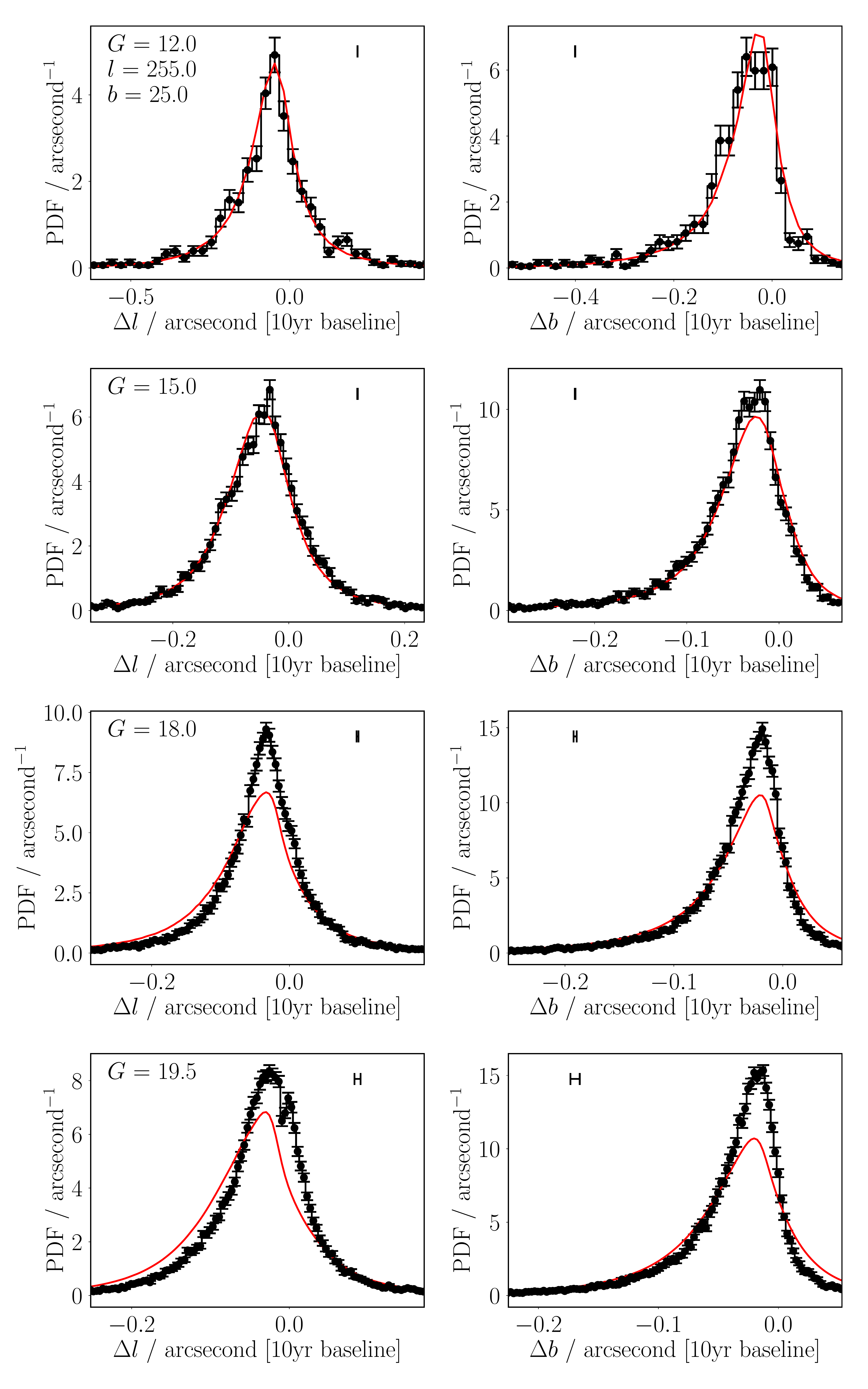}
    \caption{Distributions of proper motions for $l = 255^\circ$, $b=25^\circ$.
             Lines and symbols have the same meaning as in Figure \ref{fig:good_90}.}
    \label{fig:good_255}
\end{figure}

\begin{figure*}
    \centering
    \includegraphics[width=\textwidth]{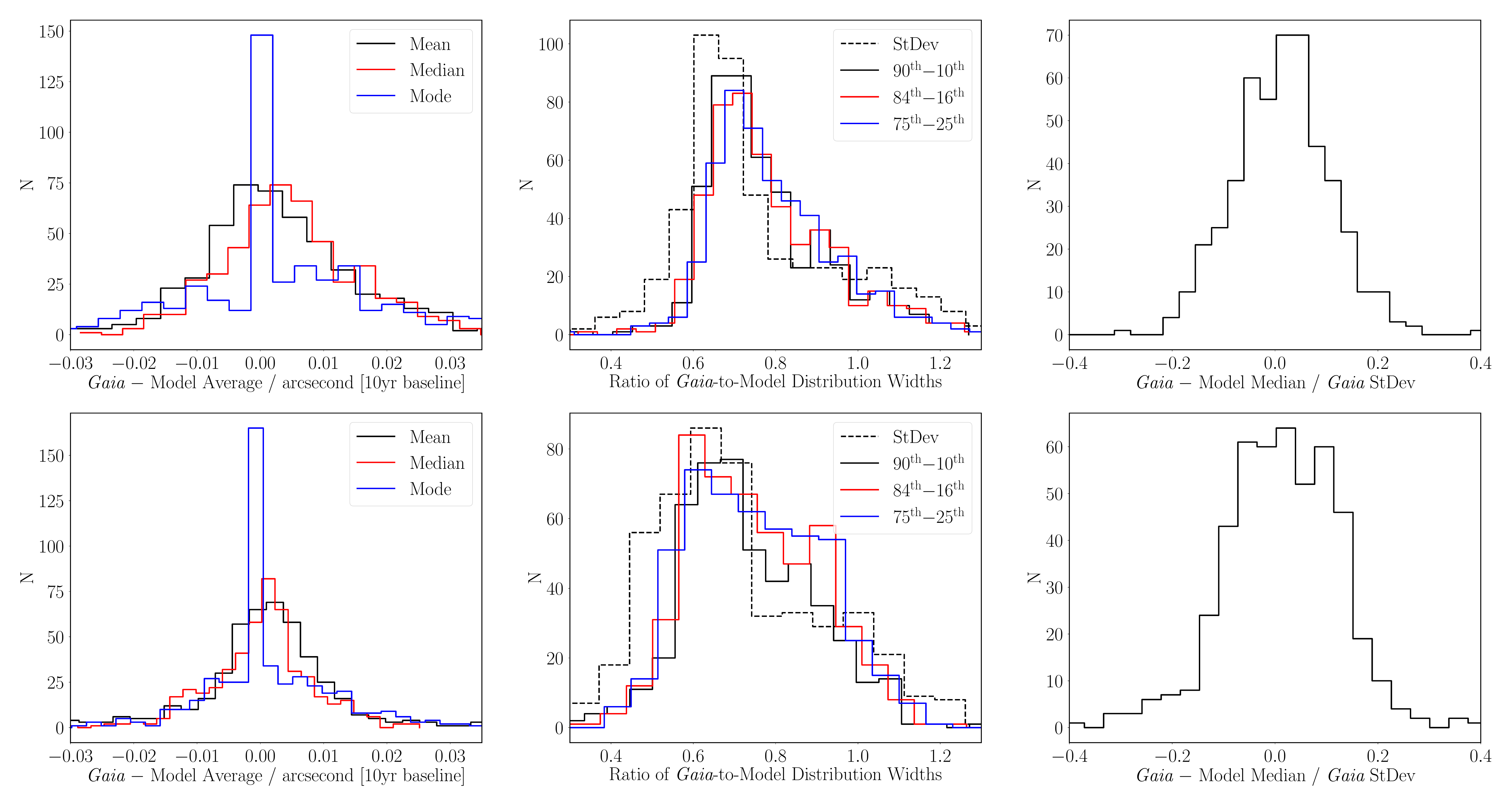}
    \caption{Comparison between \textit{Gaia} proper motions and those of our model across all sightlines and brightnesses in the Galactic plane, in Galactic longitude (top row) and latitude (bottom row).
             Left: Data-to-model average proper motion drift offsets.
             Middle: Ratio of data and model proper motion drift distribution widths.
             Right: Ratio of the median proper motion drift offset between \textit{Gaia} data and the model distribution, normalised by the standard deviation of the \textit{Gaia} proper motions.}
    \label{fig:ensemble_goodness}
\end{figure*}

\begin{figure}
    \centering
    \includegraphics[width=\columnwidth]{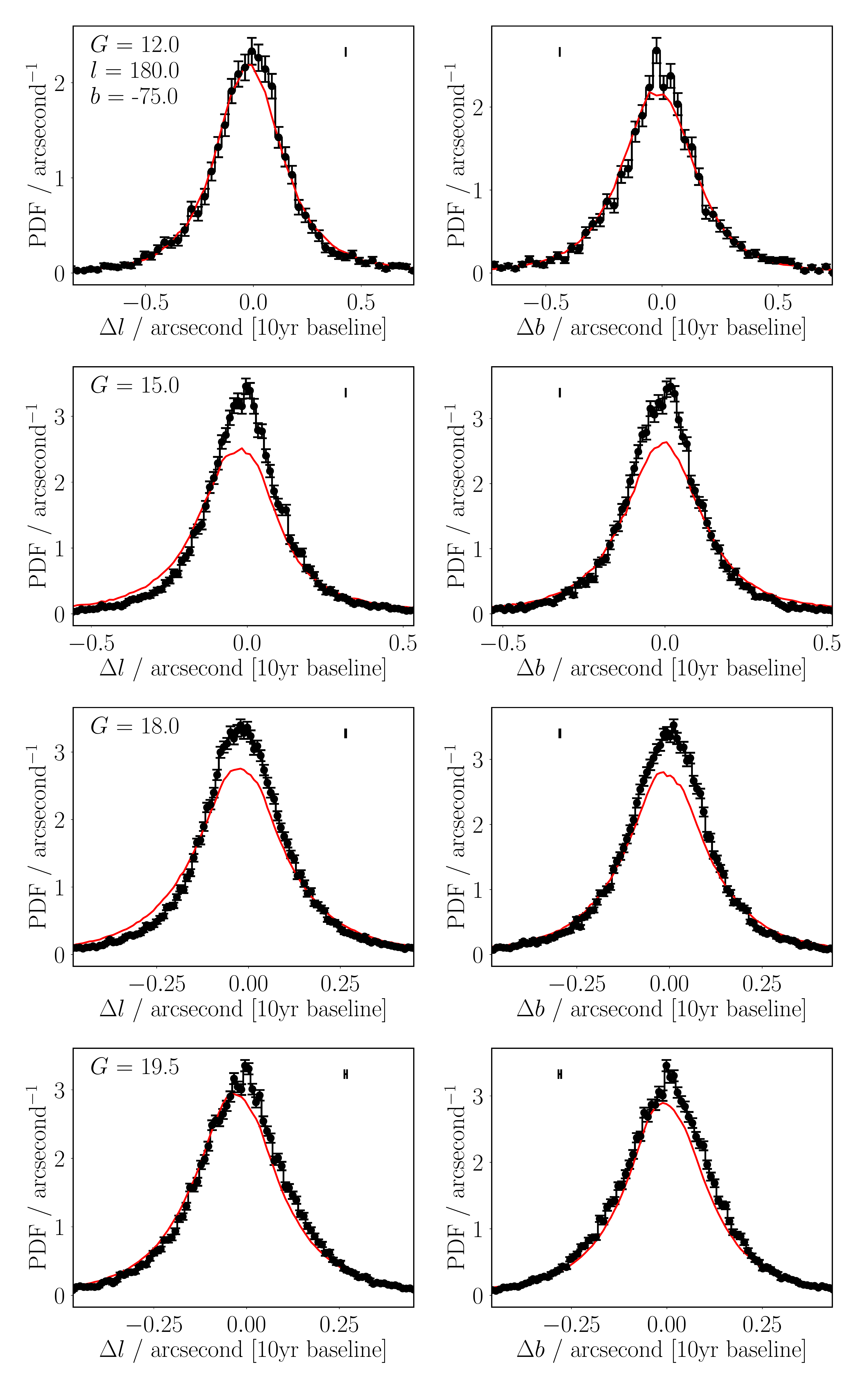}
    \caption{Distributions of proper motions for $l = 180^\circ$, $b=-75^\circ$.
             Lines and symbols have the same meaning as in Figure \ref{fig:good_90}.
             Additionally, \textit{Gaia} objects were filtered for parallaxes consistent with zero to remove extragalactic contamination.}
    \label{fig:full_third_good_high_lat}
\end{figure}

\subsection{Overall Model Shape}
The simulated proper motions are good across all sightlines and brightnesses; some examples are shown in Figures \ref{fig:good_90}-\ref{fig:good_255}.
We get good agreement in the mean proper motion, and shape of the distributions, of bulk source motions.
For most sightline-brightness combinations the agreement is quantitative, while sometimes the shapes are merely \textit{broadly} in agreement.
Disagreement in modal proper motion drift is likely largely caused by our Galactic rotation model not capturing the fine detail of Galactic potentials or inaccuracies in our asymmetric drift velocity, while distribution width issues can mostly be explained by the extrapolation of the velocity dispersion vector.
However, we stress that the model's simplicity is one of its strengths in the context of inclusion within a larger cross-match process, and that these minor differences are more than acceptable for the purpose of improving Bayesian match likelihoods.
These distributions are intended to reflect a wide range of potential positional shifts through time, rather than model any one specific proper motion.
Hence so long as the rough widths -- to within something like a factor two, which we achieve -- and mean offsets -- good to high accuracy using the Galactic rotation curve -- are modelled to reasonable accuracy, our distributions are good enough for our work, and as intended.

Figure \ref{fig:ensemble_goodness} shows some reduced statistics for the entire set of sightline-brightness combinations we tested in the Galactic plane -- $G=\{12,\,15,\,18,\,19.5\}$, $l$ in the range from $0^\circ$ to $345^\circ$ in 15 degree intervals, and $b = \{-50^\circ,\,-30^\circ,\,0^\circ,\,25^\circ,\,40^\circ\}$, as well as the Galactic north and south poles at $-90^\circ \leq b \leq -80^\circ$, $-80^\circ \leq b \leq -70^\circ$, $70^\circ \leq b \leq 80^\circ$, and $80^\circ \leq b \leq 90^\circ$.
Overall, we find that the widths of the \textit{Gaia} data are approximately 80\% that of our model (i.e. our model is too wide by 25\%) across all positions and brightnesses.
There is a roughly 10\% spread in relative widths -- middle column, Figure \ref{fig:ensemble_goodness}, cf. Figure \ref{fig:good_90}, bottom right panel, where our red model has a slightly wider wing than the histogram of the black \textit{Gaia} data.
We also see evidence for overly narrow simulated proper motion distributions ($\textit{Gaia}$-to-model ratios larger than one) along various sightlines, in approximately 8\% of cases -- but, again, get extremely good agreement along others.
These slightly-too-wide distributions are likely related to our modelled radial and vertical dependencies of the thin disc dispersion vector, as the thin disc is the dominant term at the distances our \textit{Gaia} data probe.
Additionally, as can be seen in the top row of e.g. Figure \ref{fig:good_90}, low-number statistics of brighter \textit{Gaia} stars could be interpreted as lower standard deviations, as the `real' distributions fail to probe the wings of the simulated distributions to high precision.

Relative to the standard deviation of the distributions, our biases -- the mean motion offsets -- are within $\sim10$\% two-thirds of the time, and almost always within $15-20$\%, as shown by right-hand column, Figure \ref{fig:ensemble_goodness}.
Absolutely, on a decade baseline, we find most of our mean motion offsets are within approximately 0.01 arcseconds (or 0.025 arcseconds on a 25-year baseline, as maybe be important for LSST), well within the `centroid' precision of most photometric catalogues -- left-hand column, Figure \ref{fig:ensemble_goodness}.
These results -- even where qualitative (e.g. Figure \ref{fig:good_255}, bottom-left panel) as opposed to quantitatively (e.g. Figure \ref{fig:good_255}, top-left panel) good fits -- are satisfactory, and we therefore have chosen not to over-explore the residuals, as the subtleties of the spiral arm structure of the Milky Way are outside of the scope of this work.

\subsection{Galactic Poles}
All previous examples shown (Figures \ref{fig:good_90}-\ref{fig:good_255}) were limited in Galactic latitude to $\lvert b \lvert \leq 50^\circ$, exploring primarily the proper motions of sources roughly in the Galactic plane.
However, we must also verify that our model is good at high absolute Galactic latitudes, where we are viewing sources orbiting around the Galactic center `above' us.
As shown in Figure \ref{fig:full_third_good_high_lat}, we get good agreement for the Galactic longitudinal and latitudinal proper motions, after removing objects with parallax $\pi < 0.05\,\mathrm{mas}$ ($d \gtrsim 20\,\mathrm{kpc}$) or $\pi/\sigma_\pi < 2$.    
Here, close to $b = 90^\circ$, our equations for the average rotational velocities (equations \ref{eq:uvw1}-\ref{eq:mubdef}) simplify somewhat.
First, looking straight up out of the Galactic plane, we have $d_\mathrm{ip} \approx 0$; we also have $R_c \approx R_\odot$, and hence $(R_c^2 + R_\odot^2 - d_\mathrm{ip}^2)/(2\,R_c\,R_\odot) \approx 1$, and can assume $\Theta(R_c) = \Theta_\odot$, giving $U_1 = -U_\odot$, $V_1 = -V_\odot$.
This further gives $v_d = -U_\odot \cos(l) - V_\odot \sin(l)$, $v_l = U_\odot \sin(l) - V_\odot \cos(l)$, and hence, along with a simplified $v_b = -v_d$,
\begin{align}
     \mu_{l*} &= \frac{k}{d} \left[U_\odot \sin(l) - V_\odot \cos(l)\right], \label{eq:muldef_pole}\\
     \mu_b &= \frac{k}{d} \left[U_\odot\cos(l) + V_\odot \sin(l)\right].
     \label{eq:mubdef_pole}
\end{align}
This renders the orbital motions effectively just those of the Sun, with our dispersion vector giving good shape agreement to the \textit{Gaia} data.
Overall, we see good agreement in the shape of our model and the \textit{Gaia} proper motions.

\begin{figure}
    \centering
    \includegraphics[width=\columnwidth]{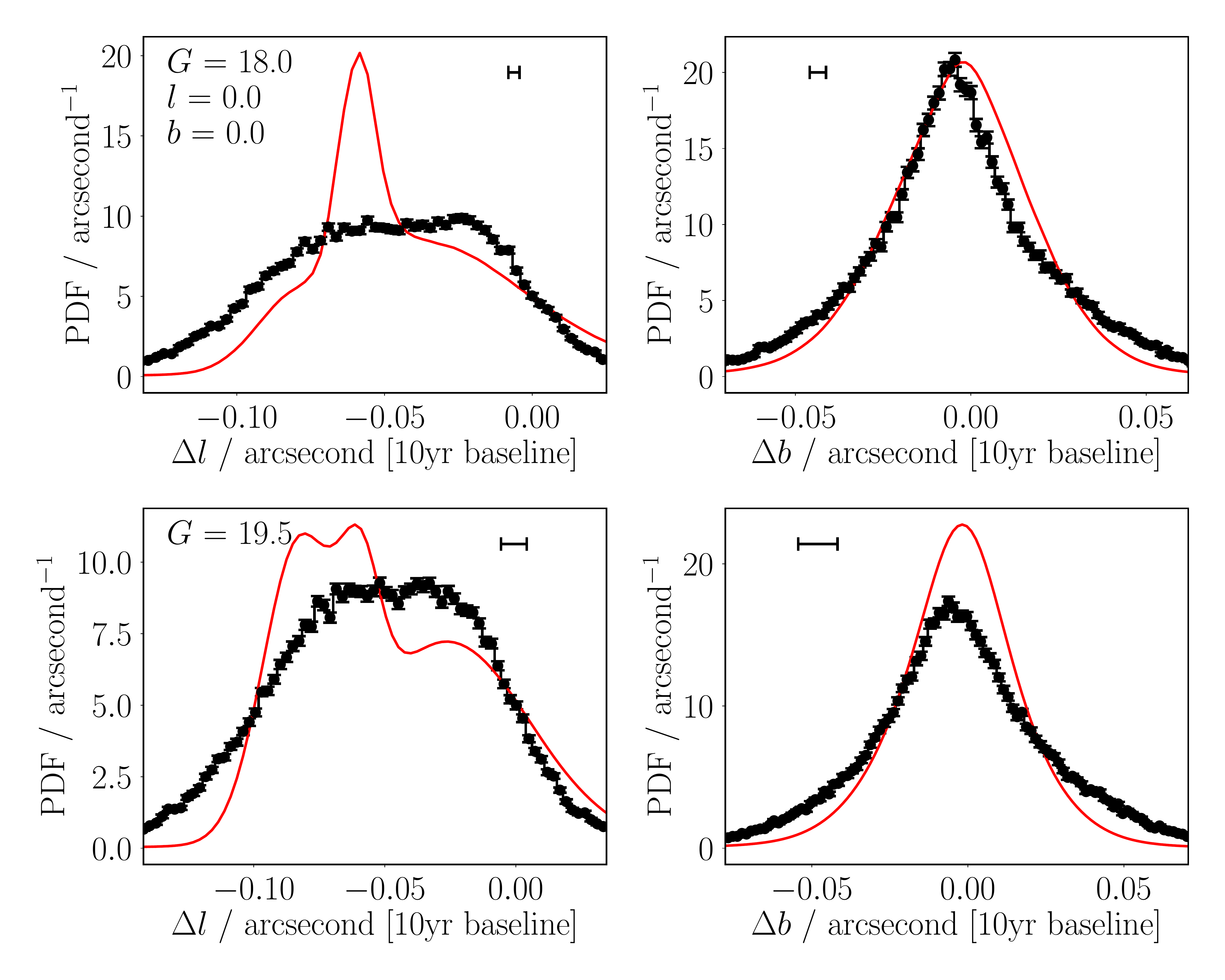}
    \caption{Distributions of proper motions for $l = 0^\circ$, $b=0^\circ$.
             Lines and symbols have the same meaning as in Figure \ref{fig:good_90}.}
    \label{fig:full_third_gc}
\end{figure}

\subsection{Widths of Distributions vs. Position and Proper Motion Precisions}
\label{sec:pmwidthvsprecision}
At this point it is worth briefly considering if, or at what brightnesses, this additional information is necessary.
Figures \ref{fig:good_90}-\ref{fig:full_third_good_high_lat} show a representative sample of Galactic sightlines and the various widths of the distributions of potential proper motions in each sightline-magnitude combination.
Overall, the widths of these distributions are approximately $0.2-0.5$ arcsecond drifts over a 10 yr baseline ($20-50\,\mathrm{mas}\,\mathrm{yr}^{-1}$) at the bright end of our tests ($G=12$), reducing to $0.1-0.3$ arcsecond drifts in 10 years ($10-30\,\mathrm{mas}\,\mathrm{yr}^{-1}$) at $G=19.5$.

The first parameter we should compare the proper motions to is the precision on an individual position.
If one or both of the positions in a given cross-match were highly uncertain, factors 10 or higher than the proper motion drift, this would dominate over the extra positional spread caused by the potential proper motion of the source.
Its inclusion would then not contribute to the determination of potential counterparts.
However, for a decade-long baseline, the spread of separations induced by unknown proper motion is at least a factor two or three higher than typical astrometric precisions, with even small motions over long enough baselines moving objects several astrometric precisions apart.
For \textit{Gaia} astrometric precisions are vastly higher; while 80\% of its sources will also have incredibly high precision proper motions even the remaining sources will have coordinate positions significantly higher than the unknown proper motion distributions.
A more typical ground-based survey, LSST should have at worse 0.07 arcsecond precision on each individual visit at $r=24$ \citep{Ivezic2019}.
While this is a factor $\approx3$ times smaller than the widths of our proper motion distributions on decade baselines, the real power of LSST lies in its repeated observations.
Depending on whether the object is in the full `Wide-Fast-Deep' (WFD) survey or in the Galactic Plane footprint, it will either be observed approximately 800 or 150 times across LSST's full survey lifetime \citep{Bianco2022}.
Hence the statistical precision on a co-added detection at $r=24$ is $0.003-0.006$ arcseconds depending on the exact number of visits.
Even including $\approx0.01\,\mathrm{arcsec}$ systematic precision \citep{Ivezic2019} this is far below the widths of our models for proper motion drift.
While those objects will likely also have proper motions after LSST DR3-4, $r=26.5$ coadded detections will have statistical astrometric precisions a factor $\sqrt{10}$ higher, $0.008-0.02$ arcseconds, still a factor 10 or more below our 10-year baseline drift spread.
It will be therefore important to take these long-baseline drifts into account for faint LSST objects.
For current-generation surveys such as SDSS, its very faintest sources have statistical positional uncertainties comparable to the tightest of our proper motion distribution widths ($\approx 0.2\,\mathrm{arcsec}$) so $r=24$ objects in SDSS may only see limited gains matching across a 10-year timespan.
Of course, the drifts increase linearly with time, and so a 15-year baseline (2015-2030, for example, in the case of SDSS-LSST) increases the potential proper motion drifts to a larger impact than astrometric precision.
Additionally, in the context of crowded field Bayesian cross-matching, even a `one-sigma' positional movement will be enough to significantly disrupt match likelihoods.

It is also useful to ask if proper motion precisions are ever comparable to the width of potential unknown proper motion.
Again, for \textit{Gaia} this is not the case due to its extremely high precision and repeated observations of all objects.
At $G=20$ the median precision on its proper motions are of order $1\,\mathrm{mas}\,\mathrm{yr}^{-1}$ \citep{Collaboration2021}.
For LSST, quoted proper motion uncertainties are also of order $1\,\mathrm{mas}\,\mathrm{yr}^{-1}$ \citep{Ivezic2019} -- but these assume $\approx 800$ visits.
Hence the stellar proper motion precisions for Galactic Plane objects will be a factor $\approx3$ higher due to the reduced number of observations within the same timeframe.
However, even $5\,\mathrm{mas}\,\mathrm{yr}^{-1}$ is $0.05\,\mathrm{arcsec}$ over a 10-year timeframe and therefore a smaller, but sizeable, fraction than the unknown proper motion distribution widths.
Right at the detection limit of proper motions with LSST it may be the case that it is preferable to not use the detected-but-unconstrained proper motions, though.
SDSS has typical limiting proper motion precisions of $5\,\mathrm{mas}\,\mathrm{yr}^{-1}$ by $r\approx20$.
Hence, like LSST, where constrained individual SDSS proper motions brighter than about 20th magnitude are likely preferable to unknown proper motions, but below this limit and the single-visit detection limit of $r=22$ unknown proper motions may be the more precise constraint.
In the IR, CatWISE \citep{Eisenhardt2020} has proper motion uncertainties of $20\,\mathrm{mas}\,\mathrm{yr}^{-1}$ at $W1\approx15$ \citep{Marocco2021} and the VVV survey \citep{2018MNRAS.474.1826S} cites uncertainties of $10\,\mathrm{mas}\,\mathrm{yr}^{-1}$ around $K_s \approx 16$; below these brightnesses the precisions on individual proper motions become comparable to or larger than the widths of typical unknown proper motion distributions.

\subsection{Missing Galactic Components}
As discussed in Section \ref{sec:veldisp}, we do not currently include a full prescription for the Galaxy.
In particular, we do not model the Galactic Bulge (or Bar).
This may have an effect at very low Galactic longitudes and latitudes.
The \textit{Gaia} data show a broader, almost flat distribution of longitudinal proper motions, where our simpler Galactic model, using the thin disc as the dominant term, has a bi-modal distribution of two narrower peaks, as shown in Figure \ref{fig:full_third_gc}, left hand panel.
It can also be seen in the data (Figure \ref{fig:full_third_gc}, right hand panel) that there is a slightly too narrow distribution of latitudinal proper motions, as compared to the \textit{Gaia} data.
This likely comes back to the minor effects of radial dependencies of the $\sigma_{rr}^2$ term, either following a Gaussian- or Rayleigh-like distribution (as discussed in Section \ref{sec:thindiskmodel}).
However, it could also be the case that our radial and vertical dispersion scalings are failing at these smaller Galactic radii, as a significant fraction of \textit{Gaia} sources ought to be sufficiently far removed from the Galactic center to be Bar or Bulge objects.
Once again, we deem these minor issues beyond the scope of this preliminary work -- the \textit{combined} bi-modal longitudinal distribution almost entirely covers the distribution of \textit{Gaia} motion drifts, to within better than a factor 1.5 or so, which is our goal.
However, we highlight the issue that the very inner few degrees of the Galactic center may suffer systematic proper motion effects due to the nature of the Galactic Bulge and Bar complexities.

We also have not modelled the Magellanic Clouds, and indeed during testing found that several of our test fields are heavily `contaminated' by sitting on the Small Magellanic Cloud (SMC) and Large Magellanic Cloud (LMC).
These extra terms, as with the Bulge, would be easy to implement; provided a relative number density of sources, with some positional distribution, and bulk and dispersal proper motion -- assuming the Magellanic Clouds are orbiting internally, and around the Milky way -- the proper motions can be modelled in much the same way with the Galactic discs and halo.
For now, we also simply urge the reader to take care when simulating sources centered on the Magellanic Clouds (SMC $l\sim300^\circ$, $b\sim-45^\circ$; LMC $l\sim280^\circ$, $b\sim-35^\circ$).

\subsection{Missing Binarity Perturbation}
Our model for motions of objects in the plane of the sky assumes all sources are single stars, subject solely to the Galactic potential.
However, half of objects are in some form of higher-order system (e.g. \citealp{Raghavan2010}) and should therefore be subject to additional on-sky motion.
It is therefore reasonable to ask whether the non-inclusion of this effect, of unresolved binary objects, would have any impact on our derived proper motions.

If the binary were equal mass, any orbital motion of the two sources around their common barycentre would completely cancel by symmetry, and show no impact on the photocentre and proper motion of the blended sources.
On the other hand, if the objects were very unequal in mass then both the barycentre and photocentre of the pair will be dominated by the larger, brighter main source, and effectively reduce to a singular object for our purposes.
An object of approximately half the mass of the primary, however, contributes very little in luminosity but significantly in astrometric effects.
Placing such an object on a worse-case orbit of approximately $10\,\mathrm{AU}$ would give an orbital period around 25 years, and a half-phase orbit on our key decade-long time interval between photometric catalogue `generations'.
In that time, the primary object would travel halfway around the orbit, appearing to move a total of $\approx6.5$AU, twice its orbital distance from the barycentre.
At a typical distance of roughly $1\,\mathrm{kpc}$ this is $6.5\,\mathrm{mas}$ or $0.54\,\mathrm{mas}\,\mathrm{yr}^{-1}$.
Such a perturbation is well below the of order $10\,\mathrm{mas}\,\mathrm{yr}^{-1}$ widths to the proper motion distributions observed for faint objects in our model.
If the object were significantly closer -- say $100\,\mathrm{pc}$ instead -- the motion effects would be 10 times higher, and comparable to the model widths.
In those cases the object would be much brighter, and likely have an individually measured proper motion or be known to be a multiple system through other means.
We therefore believe the non-inclusion of higher-order systems is justifiable at the resolution we are aiming to achieve.

\subsection{Random Positions of Sources}
As noted in Section \ref{sec:createpmdist}, the TRILEGAL simulations we use to construct our models of proper motions do not provide individual positions for simulated sources.
To overcome this, we simply uniformly distributed sources within the rectangular area we sampled our \textit{Gaia} proper motions in.
This effect may explain some small disagreements between our simulated and \textit{Gaia} proper motion distributions, as we are therefore not properly modelling any clustering, extinction effects, or other non-uniformity and correlations in the distances and positions of Galactic sources.

However, the effect on each \textit{individual} proper motion should be relatively small, as the regions in question were mostly limited to several degrees in extent, and $\cos(x+5^\circ) - \cos(x) \lesssim 0.08$, $\sin(x+5^\circ) - \sin(x) \lesssim 0.08$ over the entire Galactic longitude.
Thus our values for, e.g. the decomposition of $U_\odot$, or $\Theta$, within our proper motion equations, are of order 8\% wrong at their most extreme, in the case of a simulated patch of sky five degrees wide.
As an \textit{ensemble}, however, this assumption should be a reasonable one, and the `incorrectness' should average out, with uniformity of source distribution acceptable for small enough patches of sky, providing a statistical distribution of variations of velocity decomposition across the whole region.

\subsection{\textit{Gaia} Proper Motion Uncertainty}
To compare our ensemble proper motion distribution with the distribution of \textit{Gaia} proper motions, we included the \textit{Gaia} measurement uncertainty in our theoretical distribution of drifts.
The \textit{Gaia} data have individual uncertainties but to smooth the model with the uncertainty we had to select a single average value (the error bar included in the corners of sub-plots in e.g. Figure \ref{fig:good_90}).
A small part of the discrepancies between model and data in our analysis could therefore stem from this simplifying assumption, with no bearing on the model itself.
If the \textit{Gaia} data have a particularly broad distribution of measurement uncertainties, as they tend to at fainter magnitudes, our single uncertainty value would not produce an uncertainty-convolved motion drift distribution that reflected that of the \textit{Gaia} data.
Testing more complex treatments of \textit{Gaia} uncertainty distributions in the comparison between model and data, we found that more fully describing the non-singular value of measurement precision did produce theoretical drift distributions that better matched the expected data, but not completely.
We therefore still find a few sightlines with slight differences in central proper motion or distribution width, but perhaps 30\% of these tensions are explainable by the different measurement precisions of faint \textit{Gaia} proper motions.
Ultimately, however, as mentioned in Section \ref{sec:createpmdist}, we do not include this measurement uncertainty in the final model, just performing the convolution to compare to the \textit{Gaia} distributions more accurately.

\subsection{Comparison with the Besan\c con Model}
\label{sec:besanconcomparison}
Throughout this work we have used the TRILEGAL simulations to provide a set of theoretical distances for sources of a particular Galactic sightline and magnitude.
We could, of course, use any model of the Milky Way to achieve this, such as the Besan\c con model \citep{Robin2003,Robin2012,Robin2014,Robin2017,Czekaj2014,Bienayme2015}.
With the Besan\c con models, however, we receive simulated proper motions for the objects returned in our query, unlike with TRILEGAL.
We can use these simulated proper motions to further verify the robustness of our proper motion model; but this then raises the question of why we simply do not use these simulated proper motions for use in our cross-matches.
We will address both of these issues in the next two sections.

\subsubsection{Verifying the Accuracy of Our Model with Besan\c con}
\begin{figure}
    \centering
    \includegraphics[width=\columnwidth]{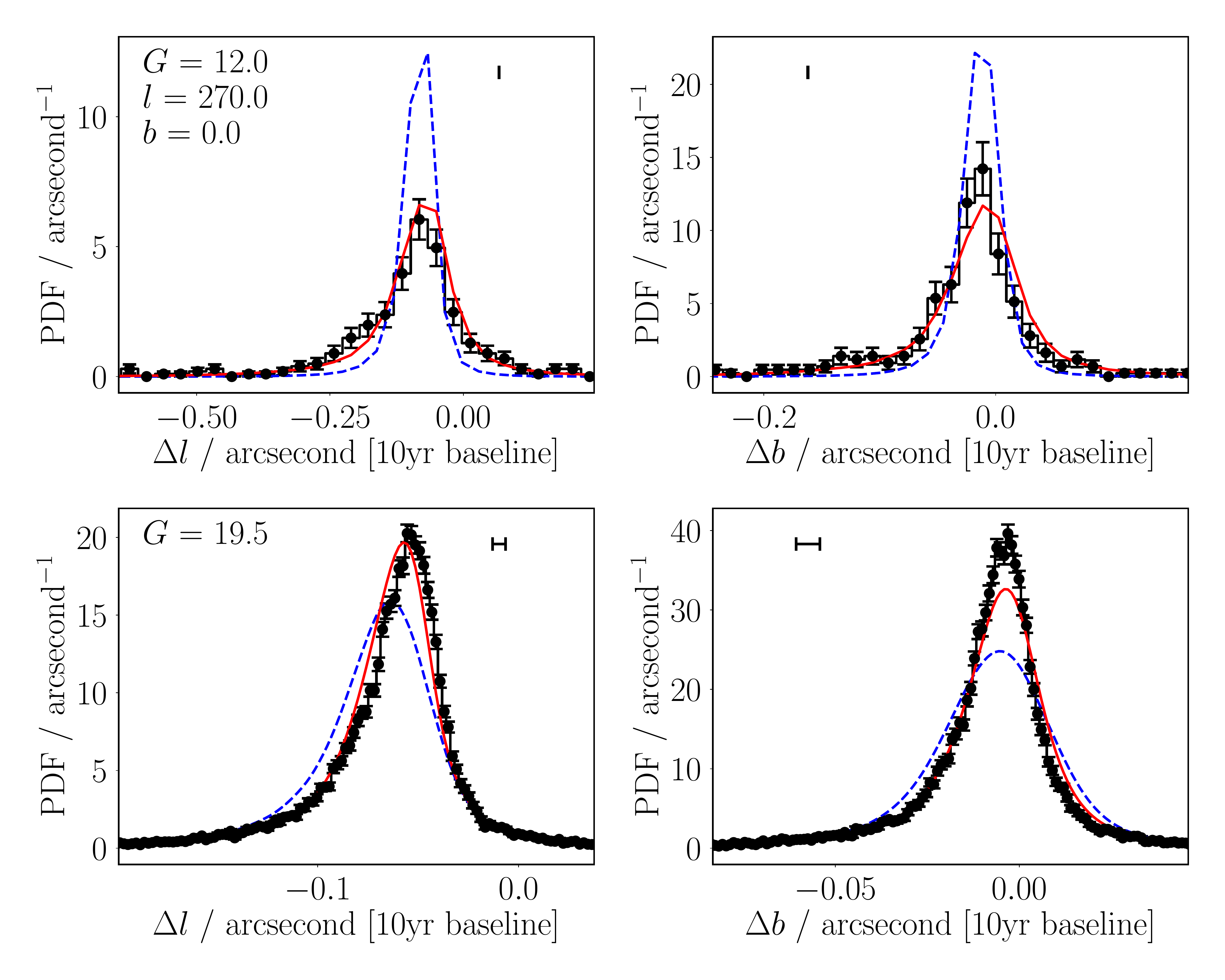}
    \caption{Distributions of proper motions for $l = 270^\circ$, $b=0^\circ$.
             Lines and symbols have the same meaning as in Figure \ref{fig:good_90}.
             In addition, the dashed blue line shows simulated Besan\c con proper motions.}
    \label{fig:besancon_270}
\end{figure}

With simulated Besan\c con proper motions, we can compare our model's statistical distribution of proper motions with those of the Galactic model.
Shown in Figure \ref{fig:besancon_270} are distributions of proper motions at $l = 270^\circ$, $b=0^\circ$ for \textit{Gaia}, our simple model for stellar velocities, and Besan\c con proper motions.
Overall, at fainter magnitudes (bottom row), both models are in agreement with the \textit{Gaia} data, with our distribution a slightly better match in Galactic latitude than the Besan\c con model.

\begin{figure}
    \centering
    \includegraphics[width=\columnwidth]{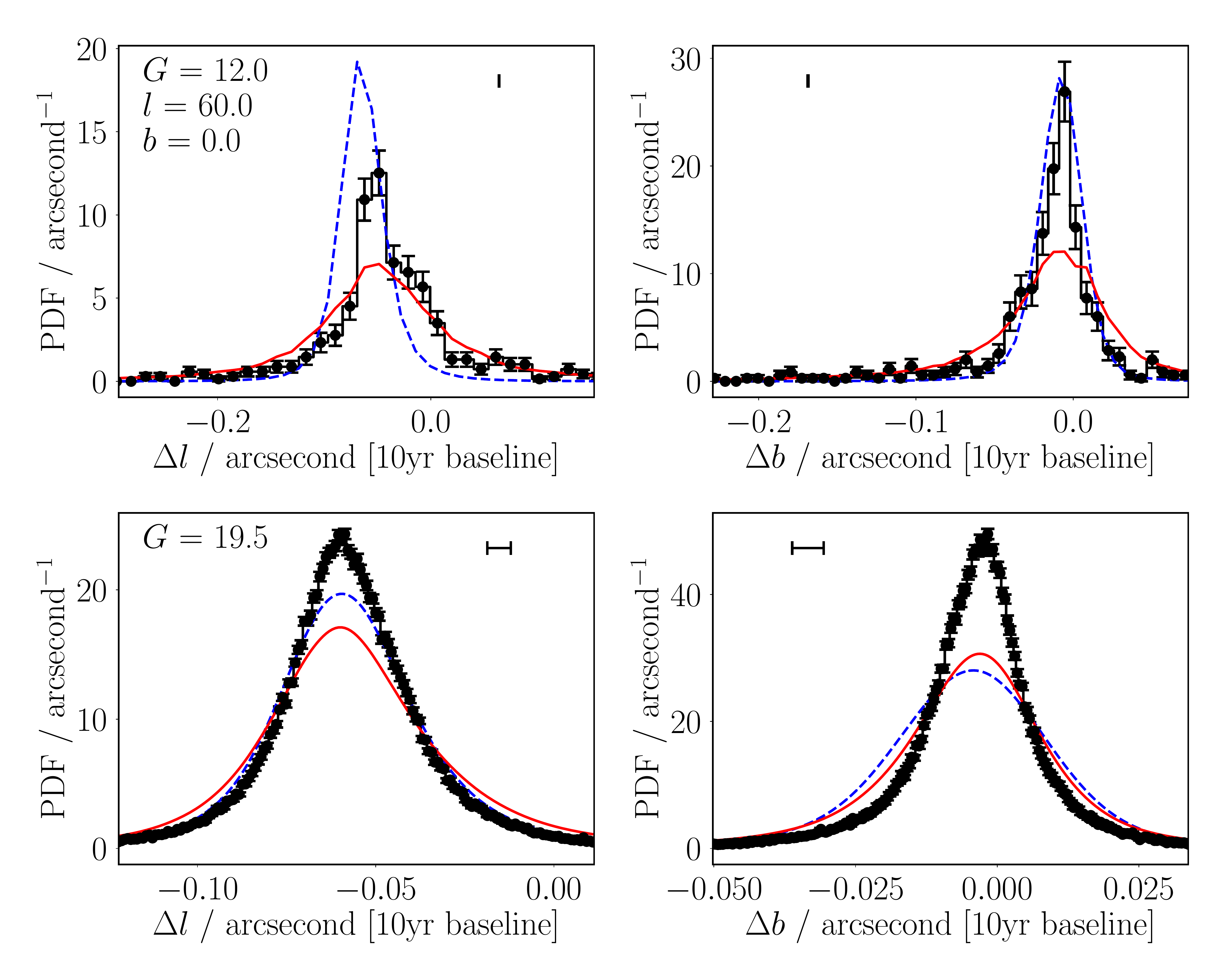}
    \caption{Distributions of proper motions for $l = 60^\circ$, $b=0^\circ$.
             Lines and symbols have the same meaning as in Figure \ref{fig:besancon_270}.}
    \label{fig:besancon_60}
\end{figure}

However, there are some sightlines within the Galaxy where our model has some mismatches to the \textit{Gaia} data -- an example sightline demonstrating this effect is shown in Figure \ref{fig:besancon_60}.
Here, at faint magnitudes ($G = 19.5$), the Besan\c con model better reproduces the Galactic longitude proper motion distribution seen with \textit{Gaia}, where our model shows a slight bias, and a distribution slightly too broad.
Neither model can reproduce the Galactic latitude \textit{Gaia} proper motions, and both look very similar in their over-broad distribution.

At bright magnitudes ($G = 12$), however, we can see that our distribution (red solid lines) much better matches the \textit{Gaia} data points than the Besan\c con simulation (blue dashed lines).
In almost all cases, $l = 60^\circ$ and $l = 270^\circ$ in Figures \ref{fig:besancon_270} and \ref{fig:besancon_60}, but more generally across multiple sightlines, the Besan\c con models are too sharp in distribution, and fail to match the \textit{Gaia} data as well as our model for proper motion.
We discuss this magnitude-dependence of the Besan\c con model fits further in Section \ref{sec:whyusebesancon}.

Here we conclude that our model matches the Besan\c con models very well, as it does the \textit{Gaia} data, and see cases where both our model and the Besan\c con model fail to match the \textit{Gaia} data perfectly. 

\subsubsection{Why Not Just Use the Besan\c con Proper Motions?}
\label{sec:whyusebesancon}
We have used TRILEGAL simulations to construct our Galactic model throughout this work, but we could have used any Galactic model.
If we had used the Besan\c con model, we would also have been provided with simulated proper motions for the objects we use for their distances in constructing our proper motions.
It is therefore reasonable to ask why we would go to the effort of using another model, if we already had a set of proper motions from which to construct a PDF of unknown proper motions.

First, as our model is broken up into separate smaller sub-models, as opposed to being wrapped in a full Galaxy model, our magnitude-to-distance relation is flexible.
As mentioned, we have been using TRILEGAL simulations to get our potential distribution of distances of sources of a given magnitude, but we could use any Galactic model.
Indeed, we do not need to use a model at all; if we instead had a known distribution of tip of the red-giant branch stars, or some other class of standard candle, we would immediately know the distance of our sources from their brightnesses.
We therefore do not necessarily need to rely on fully resolved Galactic models to provide proper motions \textit{or} distances with our simple model.
On the other hand, our options become slightly more limited if we wish to use a full Galactic model to obtain simulated proper motions in one pass, as opposed to generating more `static' distributions of object brightnesses (or distances), and using other functionality to continue on to create our final proper motion distributions, as we do here.

The second consideration is that of dimensionality; each Besan\c con source is provided with a simulated proper motion -- but only one.
Our model uses the simulated distance for each source once, but draws $N$ simulated velocities -- and hence $N$ simulated proper motions -- for each source.
We therefore much more completely sample the 3-D velocity space than any one simulation from the Besan\c con Galactic model will.
This effect can be seen in the $G=12$ panels of Figures \ref{fig:besancon_270} and \ref{fig:besancon_60}, where our model (red solid lines) much better agrees with the \textit{Gaia} proper motions, where limited sample size means the Besan\c con model is not fully populating the velocity dispersion dimensions.
At $G=19.5$, number counts have increased by a factor 100, and the velocity dispersion, having an inverse-distance component (and fainter stars being further away, on average), has reduced in size.
This reduces the effect of the lack of realisations; the Besan\c con models therefore agree much better at these fainter magnitudes than they do at bright ones.

By using each source only for its distance, as opposed to using it to sample the 4-D distance-velocity dimensionality, we much more accurately sample from the full potential proper motion distribution at bright magnitudes.
This is crucial for bright sources, being closer to the Sun on average, which have larger proper motions (cf. the $x$ axis ranges on the top and bottom rows of Figures \ref{fig:besancon_270} and \ref{fig:besancon_60}).
It is here where our constructed model has the edge on the proper motions constructed from large-scale Galactic simulations, although brighter objects are, of course, more likely to have a robustly detected proper motion from other sources.

\section{Including Proper Motions in Probabilistic Catalogue Cross-Matching}
\label{sec:includeinmatching}
No matter how you construct your proper motion distributions, it is still important to consider them in a match between two photometric catalogues of differing epochs.
The Astrometric Uncertainty Function (AUF; \citealp{2017MNRAS.468.2517W}; \citealp{2018MNRAS.481.2148W}) is the description of the belief as to a true position of the source, given its measured position.
This is typically assumed to be a Gaussian, which describes the most obvious term affecting the measured positions of sources in photometric catalogues, and hence the separation between two potential counterparts: the noise-based centroiding of the individual objects during the catalogue creation process.
The function representing the likelihood of two sources having a given separation under the assumption that they are counterparts to one another -- two detections of the same physical object -- is given by
\begin{equation}
    G(\Delta x, \Delta y) = (h_\gamma * h_\phi)(\Delta x, \Delta y)
\end{equation}
\citep{2018MNRAS.473.5570W}.
Here $\Delta x, \Delta y$ are the two-dimensional sky offsets (e.g. right ascension and declination, or Galactic longitude and latitude), $h_\gamma$ and $h_\phi$ the AUFs of the sources from the two catalogues respectively, and $(f * g)(x, y)$ denotes the convolution of two arbitrary functions $f$ and $g$ evaluated at $x$ and $y$.

As discussed by \citet{2018MNRAS.481.2148W}, the AUF $h$ can be extended with any additional terms, $h_\gamma = h_{\gamma,1} * h_{\gamma,2} * h_{\gamma,3}$ etc.
Here the additional $h_{\gamma,i}$ components, after the first noise-based centroid term, describe extra potential movement away from the `true' sky position of the source in the limit of infinite precision.
These could include, for example, stochastic processes such as the perturbation of objects due to hidden contaminants affecting the center-of-light of sources, or systematic effects like offsets of the coordinate frame of the catalogue from a common reference frame, such as the ICRS.
Whatever the effects, the point is that each source is considered individually, and has all of its $h_{\gamma, i}$ components applied to it on an isolated, per-source basis.
Proper motion, however, does not work like this; the effect of proper motion drift works on offsets between two positions, as opposed to affecting the absolute position of one source.

Thus the proper motion drift must be applied to $G$, giving, effectively $G{'} = G * h{'}_{\!\mathrm{pm}}$ (see Appendix \ref{sec:counterpartconvolvemaths} for details).
$h{'}_{\!\mathrm{pm}}$ should be calculated in the sense of mapping from oldest to youngest epoch, in units of distance; thus for $\Delta t > 0$ we have, crudely, $\Delta \delta = \mu_\delta \times \Delta t$.
Mapping from most recent to older data would have a negative $\Delta t$, but the proper motion would have to be of the opposite sign as well (being a `rewind' of the motion), and thus the sign of $\Delta \delta$ would be the same.
If a source has a purely positive proper motion distribution, such that all $\mu_{l*} > 0$ for this simulated source, then we would expect a source observed in the year J2000 to have a smaller Galactic longitude than a source observed at J2015, for example.
This convolution can be performed as any other convolution done to calculate $G$ by the convolution of all $h$ components -- e.g. either numerically, or through expression as a mixture of analytically convolvable models.

We also highlight here that while Sections \ref{sec:constructpms}-\ref{sec:assesspmmodel} detail a method for the construction of a distribution of unknown proper motions, $h{'}_{\!\mathrm{pm}}$ can be constructed through any available means.
For example, \citet{2010ApJ...719...59K} construct sets of data-driven proper motion distributions for the purpose of improving cross-matches, using available proper motions to construct priors for weighting the search for unknown proper motions between potential source counterparts.
On the other hand, the faint end of a photometric catalogue will systematically have worse precision on its measurements (see Section \ref{sec:pmwidthvsprecision}), and at some point will have detected the proper motion of an object but be unable to constrain it with high precision.
In these cases, $h{'}_{\!\mathrm{pm}}$ could very well be constructed as a Gaussian PDF with mean and covariance matrix that of the best-fit and uncertainty of the proper motion.

\subsection{Star-Galaxy Separation}
Our model for proper motions assumes the source in question is a star -- objects orbiting the Galactic center in some fashion.
However, for an all-sky catalogue cross-match we will also, at high Galactic latitudes, be matching a considerable number of galaxies.
We therefore need to model the two cases.
First, that the sources being matched are stars, and hence have the statistically modelled unknown proper motion distribution, with which we wish to `blur' out our potential match separations.
Second, they are galaxies, which have zero proper motion, being altogether too distant to have visibly moved anywhere.
We therefore have a slightly different probability of match (sources being `counterparts', under hypothesis $c$) given separation $d$, now also conditioned on the `type of source' hypothesis, which we will denote as $p(c|d, \mathcal{S})$ and $p(c|d, \mathcal{G})$ for a `star' and `galaxy' pairing respectively.

When matching, we are generally only concerned with the \textit{overall} probability of the two sources having a given sky separation under the hypothesis of their being matched, $p(d|c)$ -- this term is denoted $G$ by \citet{2018MNRAS.473.5570W}.
Note that this differs from $p(c|d)$, the probability of the two sources being counterparts given their sky separation, \cite{2018MNRAS.473.5570W}'s $g$.
$p(d|c)$ we can obtain by the marginalisation over the two hypotheses:
\begin{align}
\begin{split}
    p(d|c) &= p(d,\mathcal{S}|c) + p(d,\mathcal{G}|c) \\&=  p(d|c,\mathcal{S})P(\mathcal{S}|c) + p(d|c,\mathcal{G})P(\mathcal{G}|c)
\end{split}
\end{align}
where $P(\mathcal{S}|c)$ is the prior probability that these counterparts (with given sky positions, brightnesses, etc.) are stars (or galaxies, in the opposite case).
We work under the assumption there is no third type of object -- crudely labelling objects as `in the Milky Way' or `outside the Milky Way' -- and thus $P(\mathcal{S}|c) + P(\mathcal{G}|c) = 1$.

However, we can also ask a related but separate question: `what is the probability that these two detections are of a star, given that they are counterparts with a given separation?', which looks like
\begin{equation}
    P(\mathcal{S}|d,c) = \frac{p(d|c,\mathcal{S})P(\mathcal{S}|c)}{p(d|c)}.
\end{equation}
Here we have the likelihood of the separation given the hypothesis that the sources are counterparts and stars, multiplied by the prior chance of the sources being stars given they are counterparts, normalised by the overall chance of either a galaxy or star pair having this particular detection offset.
To calculate both $P(\mathcal{S}|d, c)$ and $p(d|c)$ we therefore need both prior and likelihood terms.

Calculating the likelihood terms $p(d|c, \mathcal{S})$ and $p(d|c, \mathcal{G})$ is relatively straightforward, simply being the convolution of the respective AUFs (containing all relevant AUF components for the two catalogues) of the sources in question.
For $p(d|c, \mathcal{G})$ this does not include any proper motion terms, as the `proper motion model' for galaxies is a static one -- mathematically, this is equivalent to the convolution of $G$ and a delta function at zero proper motion, with $f * \delta = f$ -- and hence $p(d|c, \mathcal{G}) = G$.
For $p(d|c, \mathcal{S})$, however, we wish to include the motion of Galactic sources, and hence subsequently convolve by the $h{'}_{\!\mathrm{pm}}$ PDF, describing the potential additional on-sky movement due to the epoch difference between the two sets of observations; $p(d|c,\mathcal{S}) = G{'} \equiv G * h{'}_{\!\mathrm{pm}}$.

Thus, the likelihood for our new question is easy to calculate; we are therefore left with the derivation of the prior, $P(\mathcal{S}|c)$.
The `conditioned on the fact that the sources are counterparts' aspect of the prior is tricky to implement in practice.
We wish to know, analogous to \citet{2018MNRAS.473.5570W}'s derivation of photometric likelihoods, the distribution of stars and galaxies as a function of the two bandpasses in question -- e.g. $r$ and $J$, for a match between optical and infrared data.
Thus, $P(\mathcal{S}|c)$ is really `what is the probability that these two sources are stars given that they are counterparts with magnitude limits (or dynamic ranges) in their respective bandpasses?', $P(\mathcal{S}|c, m_{\mathrm{lim,r}}, m_{\mathrm{lim, J}})$.
Due to the nature of the simulated objects -- being derived from one-sided distributions, a function of just a single magnitude in one bandpass in one of the two catalogues -- we are unable to create two-dimensional relationships between stars and galaxies in the construction of these priors.
In fact, our \textit{likelihoods}, $p(d|c,\mathcal{S})$, should implicitly assume counterparts for sources, but are built from the full distribution of sources of just a single magnitude.
Here we could have, for example, a case where sources of $J=17$ either have optical brightnesses $r=18$ or $r=25$ (being two classes of objects at differing distances, say); this distance distribution is blurred into a bimodal proper motion distribution in the IR, but one class of object is rejected if we consider the dynamic range of the optical data for an example $m_{\mathrm{lim},r} = 20$.

At present, the explicit dependency on the two-sided, magnitude-magnitude relationship between sources in our two catalogues is beyond the scope of this work, due to the nature of the outputs available from most Galactic simulations being limited to a particular set of bandpasses for a specific catalogue.
We therefore simply note here that for now, the construction of these models is one-sided -- in contrast to the cross-matching algorithms of \citet{2018MNRAS.473.5570W}, taking into account both catalogues symmetrically, in both AUF-based astrometry and photometry.
We thus sidestep this dependency by constructing our priors on star and galaxy counts on single magnitude source counts, effectively creating $P(\mathcal{S})$ and $P(\mathcal{G})$, removing the dependency on $c$ within the priors.
We can still, however, account for the dynamic range of each bandpass within its given catalogue on a per-filter basis, and hence implicitly use $P(\mathcal{S} | m_{\mathrm{lim}})$ in a practical implementation.

With this minor practical dependency removed, we conclude that with the inclusion of a distribution of unknown proper motions for Galaxy-based stars, it is possible to discriminate between stars and galaxies in photometric catalogues.
The equations
\begin{equation}
    P(\mathcal{S}|d,c) = \frac{p(d|c,\mathcal{S})P(\mathcal{S})}{p(d|c)}
\end{equation}
and
\begin{equation}
    P(\mathcal{G}|d,c) = \frac{p(d|c,\mathcal{G})P(\mathcal{G})}{p(d|c)}
\end{equation}
allow for the drift of Galactic sources with time, recovering them as non-static sources.
This is the most certain question that can be answered; stars, as shown in e.g. Figure \ref{fig:good_180}, can have a very high probability of small proper motions in certain sightlines in the Galaxy.
Thus, zero proper motion does not necessarily mean galaxy; but a combination of delta-function likelihood for Galactic proper motion and imbalanced priors at high Galactic latitudes mean that zero proper motion objects will bias towards being extragalactic.
On the other hand, if a source has a proper motion distribution which is significantly non-zero, as is the case for Galactic longitude proper motions at $l=270^\circ$, $b=0^\circ$ (Figure \ref{fig:besancon_270}), then we should see a breaking of this degeneracy.
The offset between the two sources being considered as potential counterparts should now be able to tell whether the sources are further apart than their respective AUFs would suggest -- at which point they are very likely detections of a star -- or if they have an offset compatible with their AUFs -- at which point they are very likely a galaxy.

\subsection{Inclusion of Proper Motions in the Non-Match Hypothesis}
We also note that we should also consider the proper motions within the context of non-matches, but it is easy to see that this results in a trivial case, effectively ignoring the proper motions.
For the counter hypothesis of `these sources are unrelated to one another, and separate detections of two physical sky objects', each source can have its own proper motion, based on its own statistical distribution of potential motions.
In these cases, we need to compare to the hypothesis that these sources are not related to one another given the separation between them.
This involves the double, but separate, marginalisation over all possible unknown locations \textit{and} proper motions, for both objects.
Ultimately, as the integrals are separate the proper motions do not affect the end result -- see Appendix \ref{sec:unrelatedconvolvemaths}.
This is obvious intuitively: the distribution of separations of unrelated, randomly placed objects is independent of the unknown motion history of those objects.

\section{When Are Unknown Proper Motion Distributions Needed?}
The theoretical framework for accounting for unknown proper motions presented here is relatively indifferent to the type of surveys being matched and brightnesses at which it is used.
However, practically it is more useful in some situations than others.
Hence, we summarise here some key surveys, magnitude ranges, and science cases for which the inclusion of statistical proper motions may be most crucial.

The main criterion for considering whether the inclusion of unknown proper motion distributions is important or not is the surveys being matched.
\begin{itemize}
    \item The model is most useful outside of \textit{Gaia} dynamic ranges, as such high-precision individual proper motions may be too important to ignore at brighter magnitudes. The main downside here is that \textit{Gaia} is quite a bright survey relative to the next generation of photometric catalogues, and therefore large numbers of objects won't be detected in \textit{Gaia} at all.
    \item In the case of LSST, it will also offer proper motions down to perhaps $r=24$ \citep{Ivezic2019} but cannot offer proper motions for objects not detectable in its single-visit images, and thus those objects will have to rely solely on statistical proper motions to avoid risking underestimating match probability or unnecessary false match rates.
    \item More generally, any science done in the Northern Hemisphere, where LSST has no coverage, will be unable to take advantage of the dataset -- for its increased proper motion coverage or otherwise. These cases will more likely require the falling back on unknown proper motions when outside of \textit{Gaia} or SDSS proper motion dynamic ranges ($G\approx20$ and $r\approx21$ respectively).
    \item Where proper motions are not important to the science case, and any motion drift is a nuisance parameter, it may also be preferable to avoid relying on matching to an intermediate catalogue that contains proper motions. When trying to match catalogue $A$ to catalogue $B$, we may not want to perform separate LSST-$A$ and LSST-$B$ (or \textit{Gaia}-$A$ and \textit{Gaia}-$B$, depending on your source of individual proper motions) matches, then join across common LSST (\textit{Gaia}) objects to obtain a final cross-match. In these cases, where the ability to select high-quality matches using the added-value information from a probabilistic cross-match algorithm is important, reliance on proper motion distributions may suffice to gain in other areas.
    \item With the key exceptions of CatWISE, albeit with order-of-magnitude larger uncertainties than LSST or \textit{Gaia}, and, but with much less sky coverage, VVV, most IR surveys are single-epoch, and matching longer wavelength surveys to one another therefore relies far more than optical catalogues on unknown proper motion distributions.
\end{itemize}

In terms of science cases, the main areas that benefit from including unknown proper motions are those in which proper motions are crucial but lacking by other methods.
\begin{itemize}
    \item Nearby faint objects, which LSST especially will find significant numbers of, will have appreciable on-sky motions that may not be derived as part of the survey's dataset construction due to their faint fluxes.
    \item Red objects will suffer a bias in current- and future-generation surveys such as LSST, \textit{Euclid}, and \textit{Roman} where they will systematically be less likely to have measured proper motions.
    \item Very faint transient progenitors will also suffer from a lack of known proper motions, and potentially may require matching back to a number of long-time-baseline surveys to probe progenitor characteristics.
    \item For LSST, Galactic Plane science will systematically be affected due to the much lower number of visits currently planned than in the main WFD survey \citep{Bianco2022}. Current simulated LSST precisions (e.g. \citealp{Ivezic2019}, table 3) assume WFD cadences and hence numbers of observations, but reduced visit count will lead to worse proper motion accuracies and precisions by factors of a few.
\end{itemize}

Finally, care should be taken when attempting to extrapolate reasonably uncertain, but `detected' proper motions.
In these cases it may be more advantageous to not use the best-fit value, but marginalise over all potential proper motions based on the likely more robustly determined position and brightness. Alternatively, the best-fit proper motion can be used, but `blurred' out with the detection's precision, representing the proper motion offset PDF $h{'}_{\!\mathrm{pm}}$ as a Gaussian with given mean proper motion and one-sigma uncertainty.

\section{Conclusion}
\label{sec:conclusions}
We described a model of the bulk motion of a random set of sources through the Galaxy.
The model uses the rotation curve of the Galaxy, the Solar motion, and a prescription for the random motion of sources due to e.g. their interaction history to create a statistical distribution of potential proper motions of a source at a particular set of sky coordinates and brightness.
We compared this model to \textit{Gaia} sources in various sightlines across the Galactic plane -- in the mid-plane and out of plane -- in different magnitude regimes, and to the proper motions provided by the Besan\c con Galactic model, to verify its robustness and accuracy.

Overall we find that our model matches the observed proper motions with a high degree of both accuracy and precision, and hence believe that our model is an acceptable description of the statistical proper motions of sources.
This will be invaluable when matching the next generation of deep photometric surveys to other datasets, in the regime where \textit{Gaia} cannot provide individual proper motions for sources.
Without the inclusion of unknown proper motions we could be subject to a source separation bias that will impact the number of cross-matches reported between two such catalogues.
This will be particularly crucial in the coming years in light of the revolution in Galactic studies that the Rubin Observatory's LSST will bring, where -- with its long time baseline back to previous brighter infrared surveys -- this effect has the potential to dominate a systematic search for classes of sources such as faint, red objects.

We have made a \texttt{Python} version of the model described in this paper available through the \texttt{macauff} GitHub codebase\footnote{At \href{https://github.com/Onoddil/macauff/blob/main/macauff/galactic\_proper\_motions.py}{this} URL.}.

\section*{Acknowledgements}
\label{sec:acknowledge}
TJW thanks the reviewers for their useful comments and suggestions, which much improved the manuscript.
TJW would also like to thank Sergey Koposov for useful conversations and suggestions that improved the accuracy of the model, and Tim Naylor for his helpful discussions and proofreading assistance throughout this work.
This work has been supported by STFC funding for UK participation in LSST, through grant ST/S 006117/1.
This work has made use of \textsc{Python} \citep{python3_ref}, and the \textsc{SciPy} \citep{Virtanen2020}, \textsc{NumPy} \citep{Harris2020}, \textsc{Astropy} (\citealp{2013A&A...558A..33A}; \citealp{2018AJ....156..123A}), \textsc{astroquery} \citep{Ginsburg2019}, \textsc{Matplotlib} \citep{matplotlib}, and \textsc{F2PY} \citep{f2py} \textsc{Python} modules, as well as NASA's Astrophysics Data System.

This work has made use of data from the European Space Agency (ESA) mission
{\it Gaia} (\url{https://www.cosmos.esa.int/gaia}), processed by the {\it Gaia}
Data Processing and Analysis Consortium (DPAC,
\url{https://www.cosmos.esa.int/web/gaia/dpac/consortium}). Funding for the DPAC
has been provided by national institutions, in particular the institutions
participating in the {\it Gaia} Multilateral Agreement.

\section*{Data Availability}
The datasets used in this manuscript were derived from sources in the public domain, from the \textit{Gaia} archive (\url{https://gea.esac.esa.int/archive/}), TRILEGAL (\url{http://stev.oapd.inaf.it/cgi-bin/trilegal_1.7}), and Besan\c con (\url{https://model.obs-besancon.fr/modele_home.php}).


\bibliographystyle{mnras}
\bibliography{rasti_unknownpropermotions}

\appendix

\section{Coordinate Systems}
\label{sec:coordsystems}
Here we define the coordinate systems we use in this paper, and how to transform from one to another.
We use several coordinate systems: Heliocentric Cartesian space ($x,\,y,\,z$); the observable, Heliocentric Spherical coordinate space ($d,\,l,\,b$); Heliocentric Cylindrical coordinates ($\hat{v}_d,\,\hat{v}_l,\,\hat{v}_z$); the Galactocentric Cylindrical coordinate system ($R_c,\,\phi,\,z$); the Galactocentric Spherical coordinate system ($R_s,\,\phi,\,\theta$); and the Galactocentric Cartesian coordinate system ($X,\,Y,\,Z$).

The Heliocentric Cylindrical coordinate system is defined as the radial and tangential velocity components of the in-plane stellar motions, as measured from the Sun in (and orthogonal to) the direction towards the source, as well as the orthogonal, vertical component of the motion.
Its transformation from Heliocentric Spherical coordinates is a simple rotation from $(d,\,b)$ through the angle $b$ to $(\hat{v}_d,\,\hat{v}_z)$, albeit with the caveat that the direction of rotation varies with the sign of $b$; its transformation from Galactocentric Cartesian coordinates is a rotation through longitudinal angle $l$.

The Heliocentric Cartesian coordinate system can be obtained from the Heliocentric Spherical coordinates, the observables, distance $d$, and Galactic coordinates $l$ and $b$, with
\begin{align}
    x &= d \cos(l) \cos(b) \\
    y &= d \sin(l) \cos(b) \\
    z &= d \sin(b),
\end{align}
where we have used the `right-handed' system that defines $x$ as pointing towards the Galactic center from the Sun, to $l = 0^\circ$; $y$ towards $l = 90^\circ$; and $z$ towards $b = +90^\circ$.

The Galactocentric Cartesian coordinates are a simple shift of zero-point, relative to the Heliocentric coordinates:
\begin{align}
    X &= x - R_\odot \\
    Y & = y \\
    Z &= z + z_\odot
\end{align}
with a shift of the origin up by $\simeq 8 \mathrm{kpc}$ and down $\simeq 25 \mathrm{pc}$ in the $X$ and $Z$ directions (e.g. \citealp{Juric2008}).

For the Galactocentric non-Cartesian coordinate systems, we have to define new angles, as well as two additional radii.
The radii are fairly straightforward, being based simply on the Galactocentric Cartesian coordinates.
First, the Galactocentric Cylindrical radius, being defined as the in-plane radius, is given by
\begin{equation}
    R_c = \sqrt{X^2 + Y^2}
\end{equation}
and the Galactocentric Spherical radius by
\begin{equation}
    R_s = \sqrt{X^2 + Y^2 + Z^2}.
\end{equation}

The angle $\phi$, used in both Galactocentric Cylindrical and Spherical coordinate systems, is defined as the angle around the Galaxy -- if viewed top-down, from the Galactic North Pole -- from the line running from the Sun through the Galactic center.
This angle, however, is defined as clockwise for the Galactocentric Cylindrical coordinates (and during the derivation of the Heliocentric Spherical proper motions; see Section \ref{sec:asymdrift}), but counter-clockwise for the Galactocentric Spherical coordinate system.
Equivalently, this counter-clockwise angle can be considered as being measured in the $X\,Y$ plane, from the $X$ axis towards the $Y$ axis (or $-Y$ axis, for a clockwise defined $\phi$), analagous to how $l$ is defined as the angle from the $x$ axis towards the $y$ axis.
Finally, when in Galactocentric Spherical coordinates, we could calculate $\theta$ by
\begin{equation}
    \theta = \arccos(Z / R_s),
\end{equation}
where $\theta$ is the \textit{co-latitude}, the angle as measured from the Cartesian $Z$-axis, which differs from the system defining the Galactic latitude $b$, measured from the $(x,\,y)$ plane.
We will find that we never need to consider $\phi$ or $\theta$ themselves, as they will entirely be used to define rotation.
The rotation matrices to convert from Galactocentric Spherical to Galactocentric Cylindrical coordinates, or Galactocentric Cylindrical to Heliocentric Cylindrical coordinates, along with the conversion from Galactocentric Cylindrical to Galactocentric Cartesian coordinates, are derived in Appendix \ref{sec:rotcovmatrices}.

\subsection{Rotating Covariance Matrices}
\label{sec:rotcovmatrices}
In this section we briefly outline the transformation, reflection, and rotation matrices used to convert between four coordinate systems: the Galactocentric and Heliocentric Cylindrical, and Galactocentric Cartesian and Spherical frames.
First, we need to convert from Galactocentric Cylindrical to Galactocentric Cartesian coordinates, following the rotation curve-based methodology of \citet{Mroz2019}.
Additionally, to work entirely in Sun-based radial, tangential, and vertical velocity space, we need to rotate the covariance matrices calculated from \citeauthor{Pasetto2012a}, and \citeauthor{King2015}, in Galactocentric Cylindrical and Spherical coordinates respectively, to $\hat{v}_d-\hat{v}_l-\hat{v}_z$ space.

\begin{figure}
    \centering
    \includegraphics[width=\columnwidth]{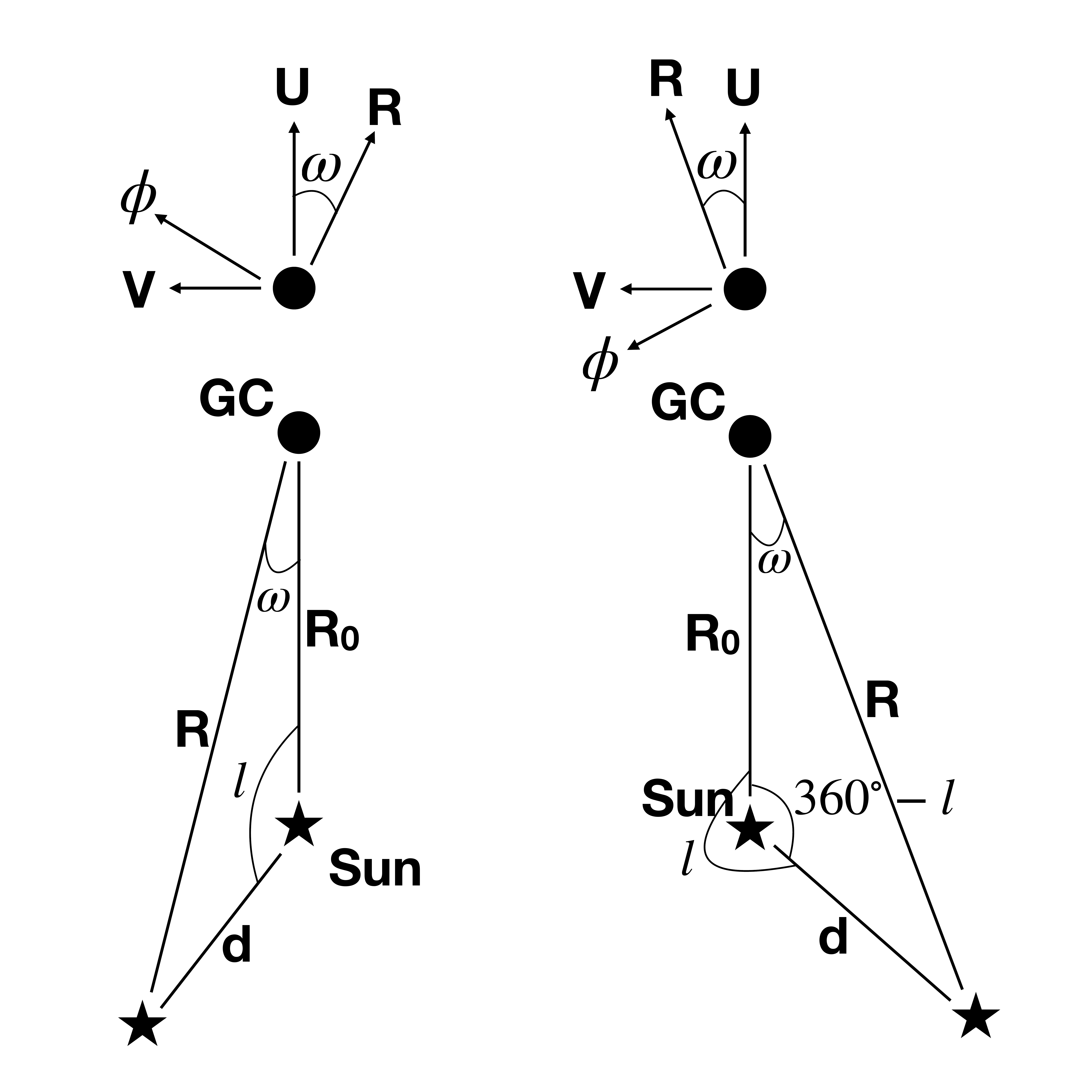}
    \caption{Schematic showing the transformation from $R-\phi$ Galactocentric Cylindrical coordinates to Galactocentric Cartesian $U-V$ coordinate system.}
    \label{fig:r-phi-U-V-setup}
\end{figure}

\subsubsection{Galactocentric Cylindrical to Galactocentric Cartesian Rotation}
\label{sec:cylcartrot}
First we need to calculate the rotation matrix describing the change from Galactocentric Cylindrical to Galactocentric Cartesian coordinates.
Starting with Figure \ref{fig:r-phi-U-V-setup}, we first consider the case of $l \leq 180^\circ$ (left-hand schematics), a counter-clockwise rotation from $R$ through angle $\omega$ to $U$.
As these are left-handed cartesian coordinate systems, this is a negative rotation, and hence the rotation matrix is
\begin{align}
\begin{split}
    \bm{\mathcal{T}}_\mathrm{CCW} = \left(\begin{matrix} \cos(\omega) & \sin(\omega) \\ -\sin(\omega) & \cos(\omega) \end{matrix}\right).
    \label{eq:left-hand-ccw-rot-mat}
\end{split}
\end{align}
We therefore need to calculate $\sin(\omega)$ and $\cos(\omega)$.
$\sin(\omega)$ can be derived using the law of sines, and is given by
\begin{equation}
    \sin(\omega) = \frac{d}{R}\sin(l),
\end{equation}
while the cosine can be calculated from its corresponding law,
\begin{equation}
    \cos(\omega) = \frac{R^2 + R_0^2 - d^2}{2\,R\,R_0}.
\end{equation}

In the $l \geq 180^\circ$ case, right-hand side of Figure \ref{fig:r-phi-U-V-setup}, we now have a clockwise rotation, which in our left-handed coordinate system is a positive rotation,
\begin{align}
\begin{split}
    \bm{\mathcal{T}}_\mathrm{CW} = \left(\begin{matrix} \cos(\omega) & -\sin(\omega) \\ \sin(\omega) & \cos(\omega) \end{matrix}\right).
    \label{eq:left-hand-cw-rot-mat}
\end{split}
\end{align}
We can, as before, calculate the sine and cosine of $\omega$:
\begin{equation}
    \sin(\omega) = \frac{d}{R}\sin(360^\circ - l) = -\frac{d}{R}\sin(l),
\end{equation}
\begin{equation}
    \cos(\omega) = \frac{R^2 + R_0^2 - d^2}{2\,R\,R_0}.
\end{equation}

We can therefore now see that the changing from positive to negative rotation in $\mathcal{T}$, which changes the sign of $\sin(\omega)$ in the rotation matrix, is correlated with a change of sign \textit{of} $\sin(\omega)$.
Thus we can simplify our matrices, giving us
\begin{align}
\begin{split}
    \bm{\mathcal{T}}_t = \left(\begin{matrix} \frac{R^2 + R_0^2 - d^2}{2\,R\,R_0} & \frac{d}{R}\sin(l) \\ -\frac{d}{R}\sin(l) & \frac{R^2 + R_0^2 - d^2}{2\,R\,R_0} \end{matrix}\right).
\end{split}
\end{align}
Expanding to the full three-dimensions of our problem, we note that the third axis is unchanged by the rotation within the plane of the Galaxy, and therefore the final axis has a trivial transformation, giving
\begin{align}
\begin{split}
    \bm{\mathcal{T}}_t = \left(\begin{matrix} \frac{R^2 + R_0^2 - d^2}{2\,R\,R_0} & \frac{d}{R}\sin(l) & 0 \\ -\frac{d}{R}\sin(l) & \frac{R^2 + R_0^2 - d^2}{2\,R\,R_0} & 0 \\ 0 & 0 & 1 \end{matrix}\right).
\end{split}
\end{align}

\begin{figure}
    \centering
    \includegraphics[width=\columnwidth]{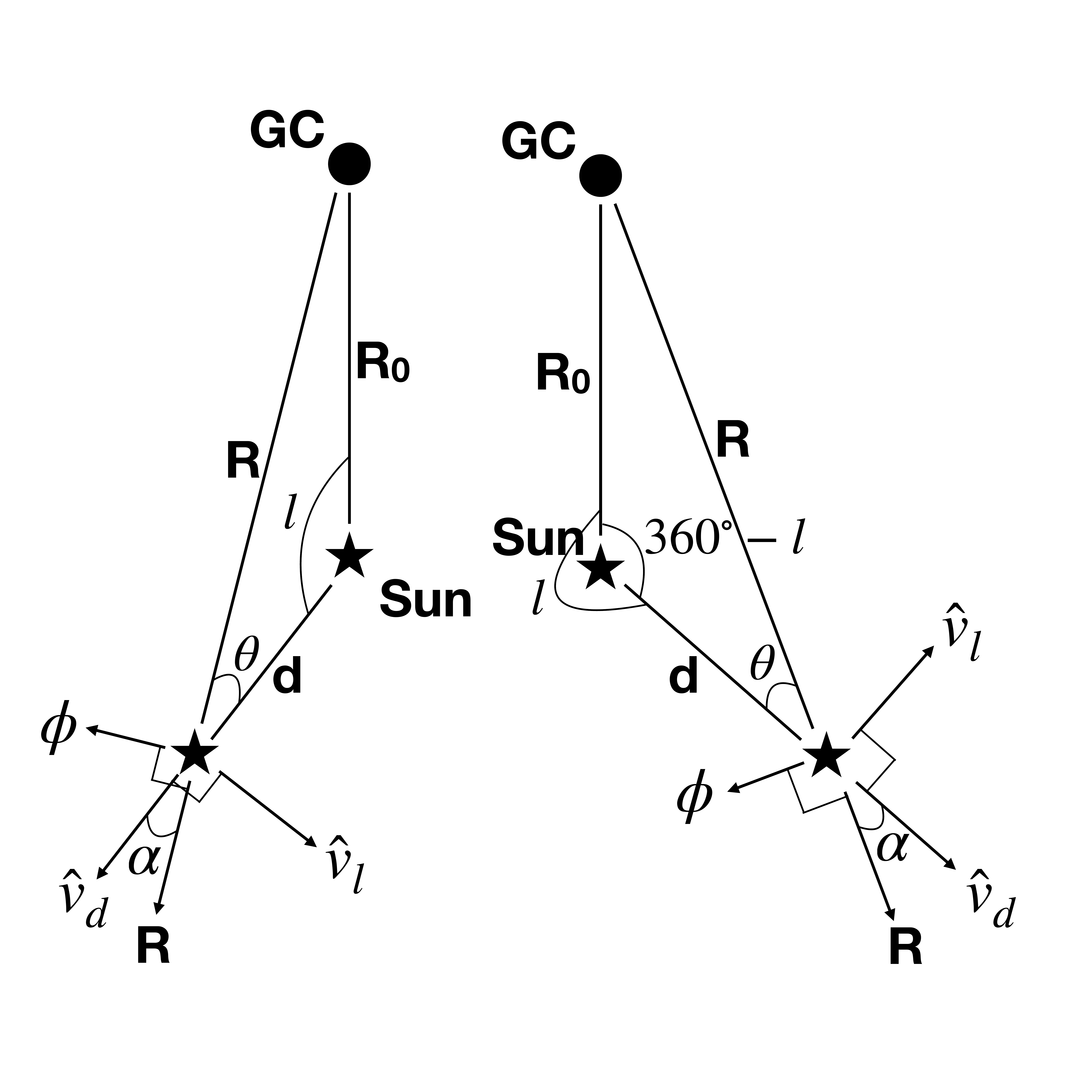}
    \caption{Schematic showing the transformation from $R-\phi$ Galactocentric coordinates to a Heliocentric $\hat{v}_d-\hat{v}_l$ coordinate system.}
    \label{fig:r-phi-vr-vt-setup}
\end{figure}

\subsubsection{Galactocentric Cylindrical to Heliocentric Cylindrical Rotation}
\label{sec:cyl_to_cyl_rot}
Here we calculate the \citeauthor{Pasetto2012a} rotation from the Galactic center-based cylindrical frame on to one centered on the Sun.
Consider the left-hand panel of Figure \ref{fig:r-phi-vr-vt-setup}; to rotate from $R-\phi$ coordinates to $\hat{v}_d-\hat{v}_l$ is a negative (clockwise) rotation -- working in the more traditional right-handed coordinate system -- through $\alpha$, as well as a mirroring around the $R$ axis (i.e. a flip of the $\phi$ axis on to the $v_l$ axis, after rotation).
Hence a rotation-then-mirror transformation matrix would look like
\begin{align}
\begin{split}
    \bm{\mathcal{T}}_\mathrm{CW} = \left(\begin{matrix} 1 & 0 \\ 0 & -1 \end{matrix}\right) \left(\begin{matrix} \cos(\alpha) & \sin(\alpha) \\ -\sin(\alpha) & \cos(\alpha) \end{matrix}\right) = \left(\begin{matrix} \cos(\alpha) & \sin(\alpha) \\ \sin(\alpha) & -\cos(\alpha) \end{matrix}\right).
\end{split}
\end{align}
As can be seen in Figure \ref{fig:r-phi-vr-vt-setup}, $\alpha = \theta$, and hence
\begin{equation}
    \cos(\alpha) = \cos(\theta) = \frac{R^2 + d^2 - R_0^2}{2 R d},
\end{equation}
\begin{equation}
    \sin(\alpha) = \sin(\theta) = \frac{R_0}{R} \sin(l).
\end{equation}

In the right-hand case of Figure \ref{fig:r-phi-vr-vt-setup}, where $l \geq 180^\circ$, we now have a counter-clockwise, positive rotation from $R$ through $\hat{v}_d$, but still have a mirror reflection. This simply changes the sign of $\sin(\alpha)$ in the rotation matrix, and hence
\begin{align}
\begin{split}
    \bm{\mathcal{T}}_\mathrm{CCW} = \left(\begin{matrix} 1 & 0 \\ 0 & -1 \end{matrix}\right) \left(\begin{matrix} \cos(\alpha) & -\sin(\alpha) \\ \sin(\alpha) & \cos(\alpha) \end{matrix}\right) = \left(\begin{matrix} \cos(\alpha) & -\sin(\alpha) \\ -\sin(\alpha) & -\cos(\alpha) \end{matrix}\right).
\end{split}
\end{align}
Again, we can consider the inner triangle of $R_0-d-R$ and calculate angles for $\alpha$ using $\theta$:
\begin{equation}
    \cos(\alpha) = \cos(\theta) = \frac{R^2 + d^2 - R_0^2}{2 R d},
\end{equation}
\begin{equation}
    \sin(\alpha) = \sin(\theta) = \frac{R_0}{R} \sin(360^\circ - l) = -\frac{R_0}{R} \sin(l).
\end{equation}

Similar to Appendix \ref{sec:cylcartrot}, we can see that no matter the direction of the rotation -- i.e. if $l \leq 180^\circ$ or $l \geq 180^\circ$ -- the sign of $\sin(\alpha)$ cancels with the sign within the $\sin(\alpha)$ elements of the transformation matrix, and hence
\begin{align}
\begin{split}
    \bm{\mathcal{T}}_\mathrm{CCW} = \bm{\mathcal{T}}_\mathrm{CW} = \left(\begin{matrix} \frac{R^2 + d^2 - R_0^2}{2 R d} & \frac{R_0}{R} \sin(l) \\ \frac{R_0}{R} \sin(l) & -\frac{R^2 + d^2 - R_0^2}{2 R d}\end{matrix}\right).
\end{split}
\end{align}

We can also now explicitly include the third axis, the vertical coordinate in our three-dimensional cylindrical reference frame, a trivial continued alignment of the $z$ axis with our $v_z$ axis, giving the final transformation matrix as
\begin{align}
\begin{split}
    \bm{\mathcal{T}}_c = \left(\begin{matrix} \frac{R^2 + d^2 - R_0^2}{2 R d} & \frac{R_0}{R} \sin(l) & 0 \\ \frac{R_0}{R} \sin(l) & -\frac{R^2 + d^2 - R_0^2}{2 R d} & 0 \\ 0 & 0 & 1 \end{matrix}\right).
\end{split}
\end{align}

\begin{figure}
    \centering
    \includegraphics[width=\columnwidth]{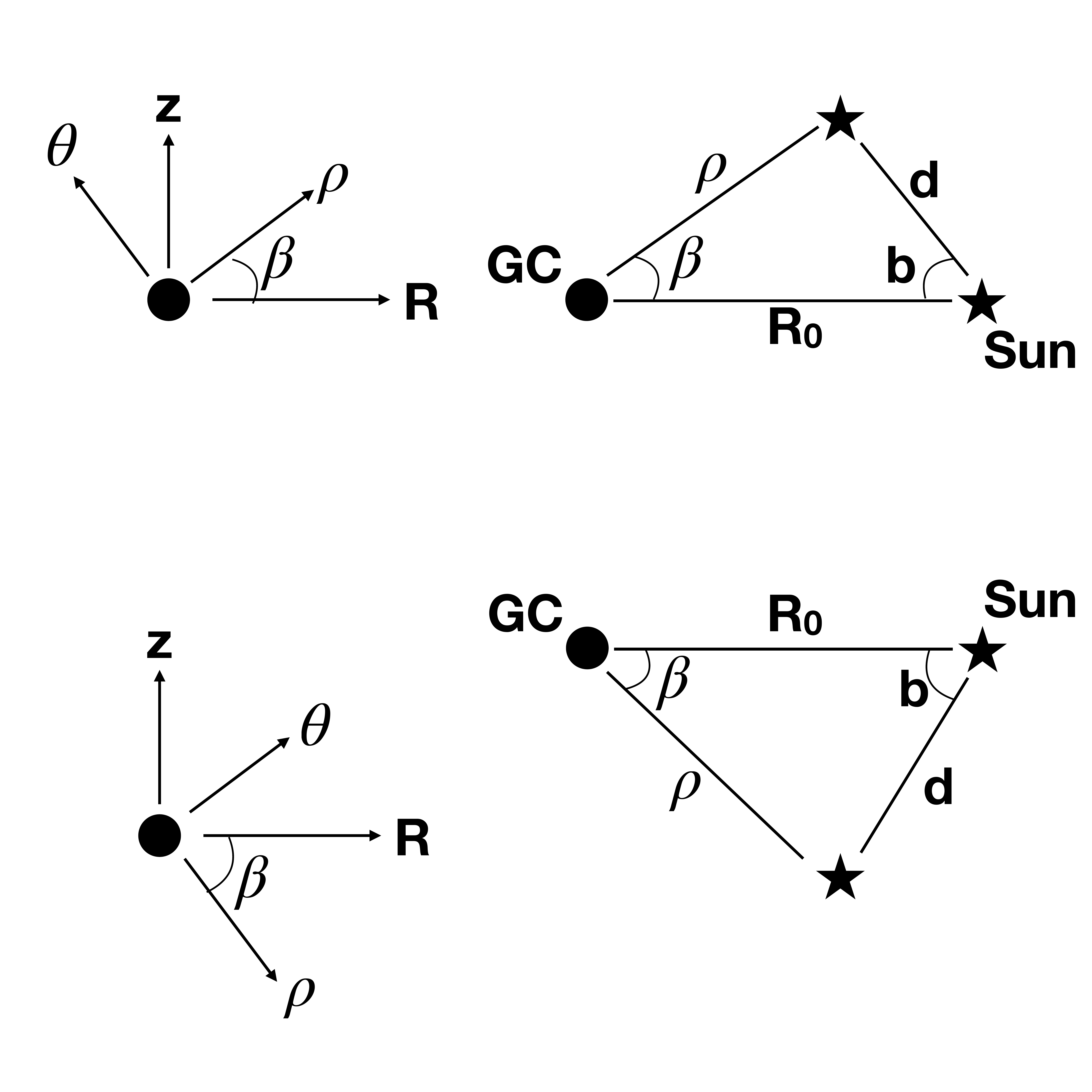}
    \caption{Schematic showing the rotation from $\rho-\theta$ Galactocentric Spherical coordinates to a Galactocentric Cylindrical $R-z$ coordinate system.}
    \label{fig:rho-theta-r-z-setup}
\end{figure}

\subsubsection{Galactocentric Spherical to Galactocentric Cylindrical Rotation}
\label{sec:sph_to_cyl_rot}
Finally, we consider the \citeauthor{King2015} rotation from a spherical reference frame into a cylindrical one, albeit still centered on the Galactic center.
To do this, we consider the frame goes from $\rho-\phi-\theta$ to $r-\phi-z$; here, ($\rho,\,\theta$) and ($R,\,z$) are both in a right-handed cartesian coordinate systems.
We therefore define our rotation matrices in the opposite sense to Section \ref{sec:cylcartrot},
\begin{align}
\begin{split}
    \bm{\mathcal{T}}_\mathrm{CW} = \left(\begin{matrix} \cos(\beta) & \sin(\beta) \\ -\sin(\beta) & \cos(\beta) \end{matrix}\right),\\
    \bm{\mathcal{T}}_\mathrm{CCW} = \left(\begin{matrix} \cos(\beta) & -\sin(\beta) \\ \sin(\beta) & \cos(\beta) \end{matrix}\right).
    \label{eq:right-hand-cw-and-ccw-rot-mat}
\end{split}
\end{align}
The upper case of Figure \ref{fig:rho-theta-r-z-setup} shows the rotation necessary for $b\geq0^\circ$, with the left hand side showing the clockwise rotation through $\beta$, and the right hand side showing a schematic of the various known distances and angles.
Here, considering a \textit{negative} -- clockwise in a right-handed frame -- rotation through $\beta$, we can calculate $\sin(\beta)$ and $\cos(\beta)$ as
\begin{align}
\begin{split}
    \sin(\beta) = \frac{d}{\rho}\sin(b),
\end{split}
\end{align}
where $\rho^2 = R_0^2 + d^2 - 2 R_0 d \cos(b)$, and
\begin{align}
\begin{split}
    \cos(\beta) = \frac{R_0^2 + \rho^2 - d^2}{2R_0\rho}.
\end{split}
\end{align}
For the lower case of Figure \ref{fig:rho-theta-r-z-setup}, $b < 0^\circ$, with a \textit{positive} rotation, $\cos(\beta) = (R_0^2 + \rho^2 - d^2)/(2 R_0 \rho)$, as previously, as the triangle is unchanged, just mirrored.
$\sin(\beta)$ is a little more complicated to derive, however, as the triangle in Figure \ref{fig:rho-theta-r-z-setup} uses $b$ as its modulus value, but it is negative in value.
Using $\lvert b \rvert$ explicitly, the law of sines gives
\begin{align}
\begin{split}
    \sin(\beta) = \frac{d}{\rho}\sin(\lvert b \rvert),
\end{split}
\end{align}
as previously.
However, if we use, as we will in practice, $b{'} = -\lvert b \rvert$, we get $\sin(b{'}) = -\sin(\lvert b \rvert)$, and hence $\sin(\beta) = -d/\rho \sin(b{'})$.

Once again, we find -- as with Appendices \ref{sec:cylcartrot} and \ref{sec:cyl_to_cyl_rot} -- that the sign of $\sin(\beta)$ cancels with the sign of the term within the rotation matrices.
Thus, for either orientation -- positive and negative Galactic latitude -- the rotation matrix from Galactocentric spherical to Galactocentric cylindrical coordinates (from the $(\rho,\,\theta)$ to $(r,\,z)$ plane) is given by
\begin{align}
    \bm{\mathcal{R}}_s &= \left(\begin{matrix} \cos(\beta) & d/\rho \sin(b)\\-d/\rho \sin(b) & \cos(\beta) \end{matrix}\right),
\end{align}
with $\cos(\beta)$ still defined consistently as before.

While we are using $\phi$ to represent the two azimuthal angles, they are defined in the opposite sense (see Section \ref{sec:coordsystems}).
We therefore need to reflect the $\phi$ axis through the $(r,\,z)$ plane, after the rotation has occurred, given by
\begin{align}
    \bm{\mathcal{R}}_\mathrm{\phi, reflect} &= \left(\begin{matrix} 1 & 0 & 0\ \\ 0 & -1 & 0 \\ 0 & 0 & 1 \end{matrix}\right).
\end{align}
Thus, our full three-dimensional transformation matrix is given by
\begin{align}
    \bm{\mathcal{R}}_s &= \left(\begin{matrix} \cos(\beta) & 0 & d/\rho \sin(b)\ \\ 0 & -1 & 0 \\ -d/\rho \sin(b) & 0 & \cos(\beta) \end{matrix}\right),
\end{align}

\section{Convolution Mathematics for Counterpart and Non-Counterpart Hypotheses}
\label{sec:convolutionmaths}
\subsection{Counterpart Likelihood Including Proper Motion}
\label{sec:counterpartconvolvemaths}
In this Appendix we detail the derivation of the inclusion of the proper motion PDF in the hypothesis that two objects are one astrophysical object given their separation.
Starting from a similar place to \citet{2018MNRAS.473.5570W}'s equation 14, we have
\begin{align}
\begin{split}
    G{'} = \iint\limits_{-\infty}^{+\infty}\!&p(\Delta u, \Delta v)\iint\limits_{-\infty}^{+\infty}\!h_\gamma(x_0 - x_\gamma, y_0 - y_\gamma)\times \\ &h_\phi(x_\phi - x_0 - \Delta u, y_\phi - y_0 - \Delta v)\,\mathrm{d}x_0\,\mathrm{d}y_0\,\mathrm{d}\Delta u\,\mathrm{d}\Delta v.
    \label{eq:gprimecounterpartfirst}
\end{split}
\end{align}
Here we have the simultaneous marginalisation over an unknown common position -- dropping the prior, $p(x_0, y_0)$ for being uniform and independent of unknown position (and proper motion), as per \citet{2018MNRAS.473.5570W} -- and a marginalisation over the PDF of all unknown proper motions drifts $p$ (here representing proper motions in the two orthogonal sky directions with $u$ and $v$).
Substituting $\Delta x = x_\phi - x_\gamma$ and $\Delta y = y_\phi - y_\gamma$ into $h_\gamma$ we get
\begin{align}
\begin{split}
    G{'} = \iint\limits_{-\infty}^{+\infty}\!&p(\Delta u, \Delta v)\iint\limits_{-\infty}^{+\infty}\!h_\gamma(x_0 - x_\phi + \Delta x, y_0 - y_\phi + \Delta y)\times \\ &h_\phi(x_\phi - x_0 - \Delta u, y_\phi - y_0 - \Delta v)\,\mathrm{d}x_0\,\mathrm{d}y_0\,\mathrm{d}\Delta u\,\mathrm{d}\Delta v.
\end{split}
\end{align}
Now we change variables from $x_0$ and $y_0$ to $x$ and $y$ via $x = x_\phi - x_0 - \Delta u$, $y = y_\phi - y_0 - \Delta v$.
This rearranges such that $x_0 - x_\phi = -\Delta u - x$, $y_0 - y_\phi = -\Delta v - y$, and thus
\begin{align}
\begin{split}
    G{'} = \iint\limits_{-\infty}^{+\infty}\!&p(\Delta u, \Delta v)\iint\limits_{-\infty}^{+\infty}\!h_\gamma(\Delta x - \Delta u - x, \Delta y - \Delta v - y)\times \\ &h_\phi(x, y)\,\mathrm{d}x\,\mathrm{d}y\,\mathrm{d}\Delta u\,\mathrm{d}\Delta v.
\end{split}
\end{align}
As per \citet{2018MNRAS.473.5570W}, we note that the inner integral is the definition of a convolution, and thus setting
\begin{align}
\begin{split}
    G&(\Delta x - \Delta u, \Delta y - \Delta v) \equiv (h_\gamma * h_\phi)(\Delta x - \Delta u, \Delta y - \Delta v) \\&= \iint\limits_{-\infty}^{+\infty}\!h_\gamma(\Delta x - \Delta u - x, \Delta y - \Delta v - y)h_\phi(x, y)\,\mathrm{d}x\,\mathrm{d}y,
\end{split}
\end{align}
we have
\begin{align}
\begin{split}
    G{'} = \iint\limits_{-\infty}^{+\infty}\!&p(\Delta u, \Delta v)G(\Delta x - \Delta u, \Delta y - \Delta v)\,\mathrm{d}\Delta u\,\mathrm{d}\Delta v.
\end{split}
\end{align}
Now it is clear that \textit{this} is itself a convolution, of $p$ and $G$, and hence we can now write
\begin{align}
\begin{split}
    G{'}&(\Delta x, \Delta y) \equiv (p * G)(\Delta x, \Delta y) \\&= \iint\limits_{-\infty}^{+\infty}\!p(\Delta u, \Delta v)G(\Delta x - \Delta u, \Delta y - \Delta v)\,\mathrm{d}\Delta u\,\mathrm{d}\Delta v.
\end{split}
\end{align}

We note that our equation \ref{eq:gprimecounterpartfirst} is of similar form to equation 5 of \citet{2010ApJ...719...59K}, with the interchange of integrals.
Here we have chosen to construct a semi-analytic simulated model for the construction of the distribution of unknown proper motions, while \citeauthor{2010ApJ...719...59K} built theirs from survey data.
However, as discussed in Section \ref{sec:includeinmatching}, we can substitute such a data-driven distribution of proper motions within our matches, using any valid distribution as $p(\Delta u, \Delta v)$ (or $h{'}_{\!\mathrm{pm}}$).

Finally, consistent with \citet{2018MNRAS.473.5570W}'s original derivation, we explicitly remind the reader that the AUFs $h$ must be defined such that $h(x, y) = h(-x, -y)$.

\subsection{Unrelated Object Likelihood Including Proper Motion}
\label{sec:unrelatedconvolvemaths}
For the case where the sources are unrelated to one another, we have a slightly different equation to that of equation \ref{eq:gprimecounterpartfirst}; something more like equation 10 of \citet{Budavari:2008aa},
\begin{align}
\begin{split}
    G{'} = \iint\limits_{-\infty}^{+\infty}\!p(\Delta u, \Delta v)&\Bigg[\iint\limits_{-\infty}^{+\infty}\!h_\gamma(x_0 - x_\gamma - \Delta u, y_0 - y_\gamma - \Delta v)\\&\,\mathrm{d}x_0\,\mathrm{d}y_0\Bigg]\,\mathrm{d}\Delta u\,\mathrm{d}\Delta v\ \times\\
           \iint\limits_{-\infty}^{+\infty}\!p(\Delta u, \Delta v)&\Bigg[\iint\limits_{-\infty}^{+\infty}\!h_\phi(x_0 - x_\phi - \Delta u, y_0 - y_\phi - \Delta v)\\&\,\mathrm{d}x_0\,\mathrm{d}y_0\Bigg]\,\mathrm{d}\Delta u\,\mathrm{d}\Delta v.
    \label{eq:gprimeunrelatedfirst}
\end{split}
\end{align}
Here, as with equation \ref{eq:gprimecounterpartfirst}, we have explicitly assumed that $p(x_0, y_0)$ is independent of both unknown position and proper motion, and thus can be removed as a factor from the equation.
As $h_\gamma$ and $h_\phi$ are normalised PDFs, the inner integral is trivially integrable to unity; but with $p$, the PDF of unknown proper motions, \textit{also} normalised, the outer integral then also evaluates to unity.
Thus we have the trivial case, for unrelated objects, that $G = 1$ and $G{'} = 1$.
In these cases, as with the counterpart hypothesis having a prior $p(x_0, y_0) = N_c$ as per \citet{2018MNRAS.473.5570W}, we can say that the equivalent priors in the `unrelated' hypothesis case are $p(x_0, y_0) = N_f$, \citet{2018MNRAS.473.5570W}'s `field' source density.
Hence, for the hypothesis of two sources being unrelated to one another and two detections of different sky objects, the PDF describing the likelihood of the objects having some separation is independent of proper motion, just as it is independent of the respective sources' AUFs.

\bsp	
\label{lastpage}
\end{document}